\documentclass[twoside,11pt]{article}
\usepackage{a4wide,cite,latexsym,amsfonts,amssymb,exscale,epsfig}
\usepackage[centertags,sumlimits,intlimits,namelimits,reqno]{amsmath}
\def\acknowledgments{\section*{Acknowledgments}}%
\def\pacs#1{\noindent {\small PACS: #1\par}}%
\def\keywords#1{\noindent {\small keywords: #1\par}}%

\usepackage{bbm}
\def\C{{\mathbbm C}}
\def\N{{\mathbbm N}}
\def\R{{\mathbbm R}}

\usepackage{amsthm}
\theoremstyle{definition}
\newtheorem{theorem}{Theorem}[section]

\newtheorem{definition}[theorem]{Definition}

\newtheorem{question}[theorem]{Question}

\newcounter{mathletter}%
\newcommand{\bmathletter}{%
  \refstepcounter{equation}%
  \setcounter{mathletter}{\value{equation}}%
  \setcounter{equation}{0}%
    \renewcommand{\theequation}{%
      \mbox{\thesection.\arabic{mathletter}\alph{equation}}}}%
\newcommand{\emathletter}{\setcounter{equation}{\value{mathletter}}}%
\newenvironment{mathletters}{\bmathletter}{\emathletter}%

\def\SO{{SO}}

\def\SU{{SU}}

\def\Spin{{Spin}}

\def\so{{\mathfrak{so}}}

\def\ie{{\sl i.e.\/}}

\def\etc{{\sl etc.\/}}
\def\cf{{\sl c.f.\/}}
\let\phi=\varphi
\let\theta=\vartheta
\let\epsilon=\varepsilon

\def\tr{\mathop{\rm tr}\nolimits}

\def\diag{\mathop{\rm diag}\nolimits}

\def\dim{\mathop{\rm dim}\nolimits}

\def\id{\mathop{\rm id}\nolimits}

\def\Hom{\mathop{\rm Hom}\nolimits}

\def\Top{\mathop{\rm TOP}\nolimits}
\def\PL{\mathop{\rm PL}\nolimits}

\def\lk{\mathop{\rm lk}\nolimits}
\def\Vert{\mathop{\rm Vert}\nolimits}
\def\sgn{\mathop{\rm sgn}\nolimits}

\let\hat=\widehat
\let\tilde=\widetilde

\newcommand{\color}[2][c]{}

\numberwithin{equation}{section}

\makeatletter

\newfont{\@aidxte}{cmsy10}
\newfont{\@aidxel}{cmsy10 scaled 1095}
\newfont{\@aidxtw}{cmsy10 scaled 1200}
\newlength\@aidxtexvi
\newlength\@aidxtexvii
\newlength\@aidxelxvi
\newlength\@aidxelxvii
\newlength\@aidxtwxvi
\newlength\@aidxtwxvii
\newcommand{\alignidx}[1]{%
  \@aidxtexvi=\fontdimen16\@aidxte
  \@aidxtexvii=\fontdimen17\@aidxte
  \@aidxelxvi=\fontdimen16\@aidxel
  \@aidxelxvii=\fontdimen17\@aidxel
  \@aidxtwxvi=\fontdimen16\@aidxtw
  \@aidxtwxvii=\fontdimen17\@aidxtw
    {\mbox{$%
    \fontdimen16\@aidxte=2.9pt
    \fontdimen17\@aidxte=2.9pt
    \fontdimen16\@aidxel=3.1pt
    \fontdimen17\@aidxel=3.1pt
    \fontdimen16\@aidxtw=3.3pt
    \fontdimen17\@aidxtw=3.3pt
    #1$}}%
    \fontdimen16\@aidxte=\@aidxtexvi
    \fontdimen17\@aidxte=\@aidxtexvii
    \fontdimen16\@aidxel=\@aidxelxvi
    \fontdimen17\@aidxel=\@aidxelxvii
    \fontdimen16\@aidxtw=\@aidxtwxvi
    \fontdimen17\@aidxtw=\@aidxtwxvii}

\makeatother

\newenvironment{myenumerate}{%
  \begin{enumerate}
  \setlength{\partopsep}{0pt}
  \setlength{\parskip}{0pt}}{\end{enumerate}}

\newenvironment{myitemize}{%
  \begin{itemize}
  \setlength{\itemsep}{0pt}
  \setlength{\parskip}{0pt}}{\end{itemize}}

\def\mycaption#1#2{%
  \begin{quote}
  \caption{\label{#1}#2}
  \end{quote}}

\def\del{\partial}
\def\emph#1{{\sl #1\/}}

\def\bra#1{\left<#1\right|}
\def\ket#1{\left|#1\right>}

\begin{document}
%
% ==============================================================================

% -----------------------------------------------------------------------------
% This is the header for article.cls.

\title{Quantum general relativity and the classification\\ of smooth manifolds}

\author{Hendryk Pfeiffer\thanks{E-mail: \texttt{H.Pfeiffer@damtp.cam.ac.uk}}}
\date{DAMTP, Wilberforce Road, Cambridge CB3 0WA, United Kingdom\\
  Emmanuel College, St Andrew's Street, Cambridge CB2 3AP, United
  Kingdom\\[1ex]
  May 17, 2004
}

\maketitle

\begin{abstract}

The gauge symmetry of classical general relativity under space-time
diffeomorphisms implies that any path integral quantization which can
be interpreted as a sum over space-time geometries, gives rise to a
formal invariant of smooth manifolds. This is an opportunity to review
results on the classification of smooth, piecewise-linear and
topological manifolds. It turns out that differential topology
distinguishes the space-time dimension $d=3+1$ from any other lower or
higher dimension and relates the sought-after path integral
quantization of general relativity in $d=3+1$ with an open problem in
topology, namely to construct non-trivial invariants of smooth
manifolds using their piecewise-linear structure. In any dimension
$d\leq 5+1$, the classification results provide us with triangulations
of space-time which are not merely approximations nor introduce any
physical cut-off, but which rather capture the full information about
smooth manifolds up to diffeomorphism. Conditions on refinements of
these triangulations reveal what replaces block-spin renormalization
group transformations in theories with dynamical geometry. The
classification results finally suggest that it is space-time dimension
rather than absence of gravitons that renders pure gravity in $d=2+1$
a `topological' theory.

\end{abstract}

\pacs{%
04.60.-m, % quantum gravity
04.60.Pp, % loops and spin foams
11.30.-j % field theory, symmetries
}
\keywords{General covariance, diffeomorphism, quantum gravity, spin foam model}

% ==============================================================================
%
\section{Introduction}
%
% ==============================================================================

Space-time diffeomorphisms form a gauge symmetry of classical general
relativity. This is an immediate consequence of the fact that the
theory can be formulated in a coordinate-free fashion and that the
classical observables are able to probe only the
coordinate-independent aspects of space-time physics, \ie\ they probe
the `geometry' of space-time which always means `geometry up to
diffeomorphism'. The classical histories of the gravitational field in
the covariant Lagrangian language are therefore the equivalence
classes of space-time geometries modulo diffeomorphism.

In the present article, we consider path integral quantizations of
classical general relativity in $d$ space-time dimensions in which the
path integral is a sum over the histories of the gravitational field.
For any given smooth $d$-dimensional space-time manifold $M$ with
boundary $\del M$, the path integral~\cite{Mi57} with suitable
boundary conditions is supposed to yield transition amplitudes between
quantum states (equation~\eqref{eq_matrixel} below). The states
correspond to wave functionals on a suitable set of configurations
that represent $3$-geometries on the boundary $\del M$. The only
background structure involved is the differentiable structure of $M$.

It is known that path integrals of this type are closely related to
Topological Quantum Field Theories (TQFTs)~\cite{At88,Wi89}. In order
to avoid possible misunderstandings of this connection right in the
beginning, we stress that in the case of general relativity in
$d=3+1$, there is no reason to require that the vector spaces [or
modules] in the axioms of~\cite{At88} be finite-dimensional [finitely
generated]. In fact, it is believed that TQFTs with finite-dimensional
vector spaces would be insufficient and unable to capture the
propagating modes of general relativity in $d=3+1$. A similar argument
is thought to apply to general relativity in $d=2+1$ if coupled to
certain matter, for example, to a scalar field.

We also stress that in the literature, the letter `T' in TQFT does not
necessarily refer to topological manifolds. In fact, the entire
formalism of TQFTs is usually set up in the framework of smooth
manifolds~\cite{At88}, and unless $d\leq 2+1$, the structure of smooth
manifolds is in general rather different from that of topological
manifolds. One should therefore distinguish smooth from topological
manifolds and relate the path integral of general relativity to a TQFT
that uses smooth manifolds. We call such a theory a $C^\infty$-QFT.
In contrast to the smooth case, we will subsequently use the term
$C^0$-QFT for a TQFT that refers to topological manifolds. One
application of the connection of general relativity with
$C^\infty$-QFTs is that the partition function which can be computed
from the path integral, forms (at least formally) an invariant of
smooth manifolds~\cite{At88,Wi89}.

Results on the classification of topological, piecewise-linear and
smooth manifolds which we review in this article, can then be used in
order to narrow down some properties of the path integral. In
dimension $d\leq 2+1$, smooth manifolds up to diffeomorphism are
already characterized by their underlying topological manifolds up to
homeomorphism. This means that in these dimensions, the $C^\infty$-QFT
of general relativity is in fact a $C^0$-QFT which makes the
colloquial assertion precise that pure general relativity in $d=2+1$
is a `topological' theory. The same is no longer true in $d=3+1$. A
given topological $4$-manifold can rather admit many inequivalent
differentiable structures so that in $d=3+1$, there is a highly
non-trivial difference between $C^\infty$-QFTs and $C^0$-QFTs. The
invariants of Donaldson~\cite{Do90} or Seiberg--Witten~\cite{Wi94} can
indeed be understood as partition functions of (generalized)
$C^\infty$-QFTs~\cite{At88} which are sharp enough to detect the
inequivalence of differentiable structures on the same underlying
topological manifold. The partition function of quantum general
relativity in $d=3+1$, if it can indeed be constructed, will offer the
same potential. This relationship is the main theme of the present
article.

The special role of space-time dimension $d=3+1$ in differential
topology is summarized by the following result.
\begin{theorem}
\label{thm_intro}
Let $M$ be a compact topological $d$-manifold, $d\in\N$ (without
boundary if $d=5$). If $M$ admits an infinite number of pairwise
inequivalent differentiable structures, then $d=4$.
\end{theorem}
This is a corollary of several theorems by various authors. We explain
in this article why this result is related to the search for a quantum
theory of general relativity.

Since in $d\geq 3+1$, smooth manifolds up to diffeomorphism are in
general no longer characterized by their underlying topological
manifolds up to homeomorphism, general relativity is no longer related
to a $C^0$-QFT. There is, however, another classification result that
is still applicable in any $d\leq 5+1$: smooth manifolds up to
diffeomorphism are characterized by their Whitehead triangulations up
to equivalence (PL-isomorphism). We explain these concepts in greater
detail below. They provide us with a very special type of
triangulations that can be used in order to discretize
smooth\footnote{This might be unexpected at first sight, but in
generic dimension $d\leq 5+1$, it is \emph{smooth} rather than
\emph{topological} manifolds that correspond to triangulations in this
way.} manifolds in a way which is not merely an approximation nor
introduces a physical cut-off, but which rather captures the full
information about the equivalence class of differentiable structures.

This suggests that the path integral of general relativity in $d\leq
5+1$ admits a discrete formulation on such triangulations. General
relativity in $d\leq 5+1$ is therefore related to what we call a
PL-QFT, \ie\ a TQFT based of piecewise-linear manifolds. In
particular, a path integral quantization of general relativity in
$d\leq 5+1$ is related to the construction of invariants of
piecewise-linear manifolds. From the classification results, we will
see that this is most interesting and in fact an unsolved problem in
topology, precisely if $d=3+1$. It is the decision to take the
diffeomorphism gauge symmetry seriously which singles out $d=3+1$ this
way.

The diffeomorphism invariance of the classical observables then
implies in the language of the triangulations that all physical
quantities computed from the path integral, are independent of which
triangulation is chosen. The discrete formulation on some particular
triangulation therefore amounts to a complete fixing of the gauge
freedom under space-time diffeomorphisms. The relevant triangulations
can furthermore be characterized by abstract combinatorial data, and
the condition of equivalence of triangulations can be stated as a
local criterion, in terms of so-called \emph{Pachner moves}. `Local'
here means that only a few neighbouring simplices of the triangulation
are involved in each step. A comparison of Pachner moves with the
block-spin or coarse graining renormalization group transformations in
Wilson's language reveals what renormalization means for theories with
dynamical geometry for which there exists no \emph{a priori}
background geometry with which we could compare the dynamical scale of
the theory.

We will finally see that the absence of propagating solutions to the
classical field equations, for example in pure general relativity in
$d=2+1$, is related
\begin{myenumerate}
\item
  \emph{neither} to the question of whether the path integral
  corresponds to a $C^0$-QFT (as opposed to a $C^\infty$-QFT),
\item
  \emph{nor} to the question of whether the vector spaces of this
  $C^0$-QFT or $C^\infty$-QFT are finite-dimensional,
\item 
  \emph{nor} to the question of whether the theory admits a
  triangulation independent discretization.
\end{myenumerate}

The present article is structured as follows. In
Section~\ref{sect_classical}, we review the classical theory and its
gauge symmetry. The formal properties of path integrals and their
connection with $C^\infty$-QFTs and manifold invariants are summarized
in Section~\ref{sect_pathint}. In Section~\ref{sect_manifolds}, we
then compile the relevant results on the classification of the various
types of manifolds and discuss their physical significance. In
Section~\ref{sect_three}, we sketch the special case of
$d=2+1$. Section~\ref{sect_mathphys} finally contains speculations on
the coincidence of open problems in physics and mathematics and on how
to narrow down the path integral in $d=3+1$. We try to make this
article self-contained by including a rather extensive Appendix which
contains all relevant definitions from topology.

% ==============================================================================
%
\section{The classical theory}
%
% ==============================================================================
\label{sect_classical}

This section contains merely review material:
Subsection~\ref{sect_manifold} on smooth manifolds,
Subsection~\ref{sect_firstorder} on the first order formulation of
general relativity, Subsection~\ref{sect_gauge} on its gauge
symmetries, Subsection~\ref{sect_difftop} on the role of differential
topology in the study of general relativity, and
Subsection~\ref{sect_diffeo} on space-time diffeomorphisms. Each of
these subsections can be safely skipped. We nevertheless include the
material here in order to fix the terminology, in particular in order
to resolve the various misunderstandings that can arise in the
discussion of diffeomorphisms in general relativity, just because
there seems to exist no standardized terminology in the literature.

%------------------------------------------------------------------------------
\subsection{Smooth manifolds}
%------------------------------------------------------------------------------
\label{sect_manifold}

\begin{figure}[t]
\begin{center}
\input{pstex/manifold.pstex_t}
\end{center}
\mycaption{fig_manifold}{%
  A manifold $M$ with two coordinate systems $(U_1,\phi_1)$ and
  $(U_2,\phi_2)$ whose patches have some overlap $U_{12}=U_1\cap U_2$,
  and the corresponding transition function $\phi_2\circ\phi_1^{-1}$.}
\end{figure}

We first review some basic facts about smooth manifolds. Detailed
definitions can be found in the Appendix.

A $d$-dimensional \emph{manifold} $M$ is a suitable topological space
which is covered by coordinate systems (Figure~\ref{fig_manifold}). A
coordinate system $(U_i,\phi_i)$ is a patch $U_i\subseteq M$ together
with a one-to-one map $\phi_i\colon U_i\to\phi_i(U_i)\subseteq\R^d$
onto some subset of the standard space $\R^d$. We can use the
coordinate maps $\phi_i$ in order to assign $d$ real coordinates
$\phi_i^\mu(p)$, $\mu=0,\ldots,d-1$, with each point $p\in U_i$. The
coordinates take values in the subset $\phi_i(U_i)\subseteq\R^d$.

As soon as two coordinate systems $(U_i,\phi_i)$, $(U_j,\phi_j)$ have
a non-empty overlap $U_{ij}=U_i\cap U_j\neq\emptyset$, we can
change the coordinates by means of the \emph{transition function}
$\phi_{ji}:=\phi_j\circ\phi_i^{-1}\colon\phi_i(U_{ij})\to\phi_j(U_{ij})$
which is a map from a subset of $\R^d$ to $\R^d$.

A scalar function $\alpha\colon M\to\R$ can be described in any of the
coordinate systems in terms of the functions
$\alpha\circ\phi_i^{-1}\colon\phi_i(U_i)\to\R$ from some subset of
$\R^d$ to $\R$.

If we require all coordinate maps $\phi_i$ to be homeomorphisms, \ie\
continuous with continuous inverse, we obtain the definition of a
\emph{topological manifold}. In this case, the transition functions
are homeomorphisms, too, and it makes sense to study continuous
functions $\alpha\colon M\to \R$ and homeomorphisms $f\colon M\to N$
between manifolds.

If we want to talk about differential equations on a space-time
manifold, we have to know in addition how to differentiate functions
and therefore have to impose additional structure. This can be
accomplished by restricting the transition functions from
homeomorphisms to some special subclass of functions. A \emph{smooth
manifold}, for example, is a topological manifold for which all
transition functions $\phi_{ji}$ and their inverses are $C^\infty$,
\ie\ all partial derivatives of all orders exist and are continuous. A
covering of $M$ with such coordinate systems is known as a
\emph{differentiable structure}. A continuous function $\alpha\colon
M\to\R$ is called \emph{smooth} if all its coordinate representations
$\alpha\circ\phi_i^{-1}\colon\phi_i(U_i)\to\R$ are
$C^\infty$-functions. A homeomorphism $f\colon M\to N$ between smooth
manifolds is called a \emph{diffeomorphism} if all coordinate
representations $\psi_j\circ f\circ
\phi_i^{-1}\colon\phi_i(U_i)\to\psi_j(V_j)$ and their inverses are
$C^\infty$. Here $(V_j,\psi_j)$ denotes the coordinate systems of $N$.

A substantial part of the present article is concerned with the
various structures one can impose on topological manifolds by
restricting the transition functions, and with the question of how to
compare these structures.

%------------------------------------------------------------------------------
\subsection{Two pictures for general relativity}
%------------------------------------------------------------------------------
\label{sect_firstorder}

For background on classical general relativity, we refer to the
textbooks, for example~\cite{MiTh73,Wa84}. We are interested in
general relativity in $d$-dimensional space-time and, for simplicity,
often restrict ourselves to pure gravity without matter. Space-time is
given by a smooth oriented $d$-manifold $M$.

\paragraph{Second order metric picture.}

We denote the (smooth) metric tensor by $g_{\mu\nu}$ from which one
can calculate the unique metric compatible and torsion-free connection
$\nabla_\mu$, its Riemann curvature tensor $R^\mu{}_{\nu\rho\sigma}$,
the Ricci tensor $R_{\mu\nu}=R^\rho{}_{\mu\rho\nu}$ and the scalar
curvature $R=g^{\mu\nu}R_{\mu\nu}$. The \emph{Einstein--Hilbert
action} (without matter) reads,
\begin{equation}
\label{eq_ehilbert}
  S[g] = \frac{1}{16\pi\,G}\int_M (R-2\Lambda)\sqrt{|\det g|}\,dx^0\wedge\cdots\wedge dx^{n-1},
\end{equation}
where $G$ is the gravitational and $\Lambda$ the cosmological
constant. Variation with respect to the metric yields the
\emph{Einstein equations} (without matter),
\begin{equation}
\label{eq_einstein}
  R_{\mu\nu} - \frac{1}{2}(R-2\Lambda)g_{\mu\nu} = 0,
\end{equation}
which are second order differential equations for the metric tensor
$g_{\mu\nu}$.

\paragraph{First order Hilbert--Palatini picture.}

We sometimes contrast this second order metric picture with the first
order Hilbert--Palatini formulation. For a recent review, see, for
example~\cite{Pe94}. We therefore choose local orthonormal basis
vectors $e_I\in T_pM$, $I=0,\ldots,d-1$, of the tangent spaces at each
point $p\in M$, \ie\ $\alignidx{e_I^\mu e_J^\nu g_{\mu\nu}=\eta_{IJ}}$
where $\eta=\diag(-1,1,\ldots,1)$ denotes the standard Lorentzian
bilinear form. The dual basis $(e^I)$ of $1$-forms $e^I=e^I_\mu
dx^\mu\in T_p^\ast M$, \ie\ $e^I(e_J)=\delta^I_J$, is called the
\emph{coframe} field. Denote by $A$ an $\SO(1,d-1)$-connection whose
curvature $2$-form we write as an $\so(1,d-1)$-valued $2$-form
$F^I{}_J=dA^I{}_J+A^I{}_K\wedge A^K{}_J$ on $M$. Classical general
relativity can be formulated as a first order theory with the
\emph{Hilbert--Palatini action},
\begin{equation}
\label{eq_firstorder}
   S[A,e] = \frac{1}{16\,(d-2)!\,\pi\,G}\int_M
    \epsilon_{IJK_1\cdots K_{d-2}}\,e^{K_1}\wedge\cdots\wedge e^{K_{d-2}}\wedge
    \Bigl(F^{IJ} - \frac{2\Lambda}{d(d-1)}\,e^I\wedge e^J\Bigr),
\end{equation}
where $F^{IJ}=F^I{}_K\eta^{KJ}$.
Variation with respect to the connection $A$ and the cotetrad $e$
yields the field equations,
\begin{mathletters}
\label{eq_field}
\begin{eqnarray}
  0 &=& \epsilon_{IJPK_1\cdots K_{d-3}}\,e^{K_1}\wedge\cdots\wedge e^{K_{d-3}}\wedge
        \Bigl(F^{IJ}-\frac{2\Lambda}{(d-1)(d-2)}\,e^I\wedge e^J\Bigr),\\
  0 &=& \bigl(\delta^I_L\,d + 2\,A^I{}_L\wedge\bigr)\,
        \epsilon^{LJ}{}_{K_1\cdots K_{d-2}}\,e^{K_1}\wedge\cdots\wedge e^{K_{d-2}},
\end{eqnarray}%
\end{mathletters}%
for all $I,J,P$. These are coupled first order differential equations
for $A$ and $e$. Whenever the cotetrad is non-degenerate, the first of
these equations is equivalent to~\eqref{eq_einstein} for the metric
tensor
\begin{equation}
\label{eq_metric}
  \alignidx{g_{\mu\nu}=e^I_\mu e^J_\nu\eta_{IJ}},
\end{equation}
while the second equation states that the connection is
torsion-free. Note that any $\SO(1,d-1)$-connection is always metric
compatible with respect to~\eqref{eq_metric}.

One reason for introducing the Hilbert--Palatini formulation is to
demonstrate that we can easily adopt a point of view in which the
metric tensor is not a fundamental field of the classical theory. This
illustrates once more that a generic smooth manifold $M$ is the only
input of the theory. The fields $A$ and $e$ and, as a consequence, the
metric $g$ are determined by the dynamics of the theory.

%------------------------------------------------------------------------------
\subsection{Gauge symmetries}
%------------------------------------------------------------------------------
\label{sect_gauge}

Let us review the gauge symmetries of the classical theory in the
first order picture.

\paragraph{Local Lorentz symmetry.}

The choice of local orthonormal bases $(e_I)$ is unique only up to a
local Lorentz transformation. Such transformations can be expressed in
a coordinate system $U\subseteq M$, $\phi\colon U\to\R^d$ by some
smooth Lorentz-group valued function $\Lambda\colon
U\to\SO(1,d-1)$. We write $x=\phi(p)$, $p\in U$, and denote the old
variables by $(A,e)$ and the transformed ones by $(\tilde A,\tilde
e)$,
\begin{mathletters}
\label{eq_localgauge}
\begin{eqnarray}
\label{eq_locallorentz}
  \tilde e_\mu^I(x) &=& \Lambda^I{}_J(x) e_\mu^J(x),\\
  \tilde A_\mu{}^I{}_J(x) &=& 
    \Lambda^I{}_K(x) A_\mu{}^K{}_L(x) \Lambda_J{}^L(x) +
    \Lambda^I{}_K(x) \frac{\del}{\del x^\mu} \Lambda_J{}^K(x),
\end{eqnarray}%
\end{mathletters}%
where we write $\Lambda_J{}^K=\eta_{JI}\eta^{KL}\Lambda^I{}_L$. The
metric~\eqref{eq_metric} is obviously invariant
under~\eqref{eq_locallorentz}.

\paragraph{Space-time diffeomorphisms.}

Any two geometries $(M,g)$ and $(M,g^\prime)$ of $M$ are physically
identical as soon as they are related by a space-time diffeomorphism
$f\colon M\to M$, \ie\ $g^\prime=f^\ast g$.

Let us describe the action of some diffeomorphism $f\colon M\to M$ on
the various fields in detail. Consider a smooth function $\alpha$, a
smooth vector field $X=X^\mu\del_\mu$, a smooth $1$-form
$\omega=\omega_\mu\,dx^\mu$, and the coframe field
$e^I_\mu\,dx^\mu$. For $p\in M$, choose a coordinate system $(U,\phi)$
such that $p\in U$ and write $x=\phi(p)$. The diffeomorphism acts in
coordinates as follows,
\begin{mathletters}
\label{eq_diffaction}
\begin{eqnarray}
  (f^\ast\alpha)(x)       &=& \alpha(f(x)),\\
  {(f^\ast X)}^\mu(x)     &=& {(Df^{-1}(f(x)))}^\mu{}_\nu\, X^\nu(f(x)),\\
  {(f^\ast\omega)}_\mu(x) &=& \omega_\nu(f(x))\,{(Df(x))}^\nu{}_\mu,\\
  {(f^\ast e)}^I_\mu(x)   &=& e^I_\nu(f(x))\,{(Df(x))}^\nu{}_\mu,\\
  {(f^\ast A)}_\mu{}^I{}_J(x) &=& A_\nu{}^I{}_J(f(x))\,{(Df(x))}^\nu{}_\mu,
\end{eqnarray}%
\end{mathletters}%
where ${(Df(x))}^\mu{}_\nu:=\del f^\mu(x)/\del x^\nu$ denotes the
Jacobi matrix of $f$, written in coordinates, too. These rules also
determine the action of the diffeomorphism on higher rank
tensors\footnote{We have defined the diffeomorphism action on
$e^I_\mu$ so that it acts only on the cotangent index $\mu$, but not
on the internal index $I$.}.

%------------------------------------------------------------------------------
\subsection{Differential topology versus differential geometry}
%------------------------------------------------------------------------------
\label{sect_difftop}

The \emph{problem of classical general relativity} can be summarized
as follows.

\paragraph{Second order picture.}

Given a smooth oriented $d$-manifold $M$ with boundary $\del M$, find
a smooth metric tensor $g_{\mu\nu}$ that satisfies the Einstein
equations~\eqref{eq_einstein} in the interior of $M$ and suitable
boundary conditions on $\del M$. Study existence and uniqueness of the
solutions. Any two solutions $\alignidx{g_{\mu\nu},g^\prime_{\mu\nu}}$
that are related by a space-time diffeomorphism $f\colon M\to M$, \ie\
$g^\prime=f^\ast g$, are physically identical.

\paragraph{First order picture.}

Given a smooth oriented $d$-manifold $M$ with boundary $\del M$, find
a smooth $\SO(1,d-1)$-connection $A$ and a smooth non-degenerate
coframe field $e$ that satisfy the first order field
equations~\eqref{eq_field} in the interior of $M$ and suitable
boundary conditions on $\del M$. Study existence and uniqueness of the
solutions. Any two solutions $(A,e)$, $(A^\prime,e^\prime)$ that are
related by a local Lorentz transformation~\eqref{eq_localgauge} or by
a space-time diffeomorphism~\eqref{eq_diffaction}, are physically
identical.

\paragraph{Interpretation.}

We stress that the input for classical general relativity is a smooth
manifold, \ie\ a topological manifold with a differentiable
structure. We are therefore in the realm of differential
\emph{topology}. This has to be contrasted, for example, with the
non-generally relativistic treatment of Yang--Mills theory in an
\emph{a priori} fixed space-time geometry. Such a background geometry
is described by a Riemannian manifold $(M,g)$, \ie\ by a smooth
manifold $M$ with a metric tensor $g_{\mu\nu}$, which renders such a
theory a problem of differential \emph{geometry} rather than
differential \emph{topology}.

\paragraph{Boundary conditions.}

We do not comment on how the boundary and/or initial conditions affect
the existence and uniqueness of the classical solutions. For
simplicity, we use the following specification of boundary data in the
first order formulation. In the variational principle, we fix the
connection $A|_{\del M}$ at the boundary $\del M$ and assume that the
variations $\delta A$ and $\delta e$ are supported only in the
interior of $M$. The field equations~\eqref{eq_field} in the interior
of $M$ can then be derived without additional boundary terms for the
action.

This choice is made entirely for convenience. It is known to be
satisfactory in the toy model of pure gravity in $d=2+1$, but might
have to be revised in order to find the correct path integral in
$d=3+1$.

%------------------------------------------------------------------------------
\subsection{Gauge symmetries and quantization}
%------------------------------------------------------------------------------
\label{sect_diffeo}

Assuming that there exists a quantization of general relativity, will
the quantum theory respect the classical gauge symmetries, or not? In
order to get some more insight into this question, let us recall why
general relativity possesses such symmetries in the first place. The
answer is quite standard~\cite{Wa84}, but nevertheless worth spelling
out in detail, in particular because this symmetry plays an important
role in the remainder of the present article.

The entire framework of smooth manifolds has been set up in order to
describe physical quantities by mathematical constructions that have a
meaning independently of the coordinate systems which we choose in
order to represent them.

Consider, for example, some physical field which is given by a smooth
scalar function $\alpha\colon M\to\R$. There are various possible
coordinate systems $(U_i,\phi_i)$ in order to represent this scalar
field. The reader may think of flat space-time with Cartesian or polar
coordinates which give rise to coordinate representations of the
field, $\alpha\circ\phi_j^{-1}\colon\phi(U_j)\to\R$, and whose
transition functions
$\phi_{ji}=\phi_j\circ\phi_i^{-1}\colon\phi_i(U_i\cap
U_j)\subseteq\R^d\to\phi_j(U_i\cap U_j)\subseteq\R^d$ prescribe how to
change the coordinates.

One might now be tempted to think that the physical reality is
described literally by the set $M$ with its points $p,q\in M$ which
would symbolize space-time events. Starting from this `reality', one
would then construct various coordinate systems $\phi_i\colon
U_i\subseteq M\to\R^d$ in order to represent this `reality' in terms
of coordinates. The transition functions guarantee that iterated
coordinate changes always give consistent results.

Certainly, the collection of all the ranges $\phi_j(U_j)\subseteq\R^d$
of the coordinate systems together with the transition functions
$\phi_j\circ\phi_i^{-1}$ and the coordinate representations
$\alpha\circ\phi_j^{-1}$ of the scalar function constitute the maximum
information about our `reality' that can be written down by the
experimenters (who always use coordinate systems). This raises the
question of whether, conversely, we can reconstruct $M$ together with
its points $p,q\in M$ and the function $\alpha\colon M\to\R$ from such
a collection of coordinate ranges, transition functions and coordinate
representations. The well-known answer to this question is
`no'~\cite{Wa84}. One can reconstruct\footnote{A closely related
result in mathematics is the fact that fibre-bundles can be
reconstructed from their transition functions, but only up to bundle
automorphisms, \ie\ compositions of local frame transformations and
diffeomorphisms.} $M$ and $\alpha$ \emph{only up to
diffeomorphism}. The actual reality is therefore not given literally
by the set $M$ and the function $\alpha\colon M\to\R$, but rather by
equivalence classes $[(M,\alpha)]$ modulo diffeomorphism. Here
$(M,\alpha)$ and $(M^\prime,\alpha^\prime)$ are considered equivalent
if and only if there exists a diffeomorphism $f\colon M\to M^\prime$
such that $\alpha^\prime=f^\ast\alpha$. We have called this a
\emph{gauge symmetry} simply because there are several different
mathematical ways $(M,\alpha)$, $(M^\prime,\alpha^\prime)$, $\ldots$
of specifying a single physical history.

If we now attempt to quantize such a theory, we have to remember that
the classical configurations (histories) are the equivalence classes
$[(M,\alpha)]$ rather than the particular representatives
$(M,\alpha)$. The above argument indicates that the quantum theory
should not violate this type of gauge symmetry because otherwise the
outcome of quantum experiments would depend, very roughly speaking, on
whether the classical observer who performs the measurement, uses
Cartesian or polar coordinates in order to write down the
observations.

As classical general relativity is formulated in the language of
smooth manifolds, it is automatically compatible with the action of
space-time diffeomorphisms\footnote{Strictly speaking, the generic
transformations are diffeomorphisms $f\colon M\to M^\prime$, but there
are various diffeomorphisms $f_1,f_2,\ldots\colon M\to M^\prime$
between the same pair of manifolds, and we can use the first, $f_1$,
in order to identify $M^\prime\equiv M$ and then obtain a
diffeomorphism $M\to M$ from the second, \etc. One can therefore say
that the symmetry is given by space-time diffeomorphisms $M\to M$.}
$f\colon M\to M$ and therefore well-defined as a theory on the
equivalence classes of geometries. It is thus safe to choose a
particular representative of the geometry whenever convenient and to
perform the relevant calculations for the representative.

When one tries to apply a quantization procedure to the theory, the
same is no longer obvious. For many quantization schemes, one is
forced to choose representatives and to write down all the physical
fields as particular functions $\alpha\colon M\to\R$, \ie\ as fields
\emph{on} space-time. The study of the equivalence classes and the
central question of whether the quantization scheme was indeed
compatible with space-time diffeomorphisms, are often postponed.

In the subsequent sections, we insist on working with the equivalence
classes. This leads us in particular to the question of when two given
smooth manifolds are diffeomorphic.

What is the difference between general relativity and other theories
such as the non-generally relativistic treatment of Yang--Mills
theory? That theory, too, can be formulated in a coordinate-free
fashion and therefore shares the same type of gauge symmetry, \ie\ any
two representations $(M,g,A,\psi)$ and
$(M^\prime,g^\prime,A^\prime,\psi^\prime)$ are physically equivalent
if and only if they are related by a diffeomorphism $f\colon M\to
M^\prime$, \ie\ $g^\prime=f^\ast g$, $A^\prime = f^\ast A$, and
$\psi^\prime= f^\ast\psi$. Here $(M,g)$ is the underlying Riemannian
manifold and $A$ and $\psi$ denote the additional fields, in the
Yang--Mills case a connection and some charged fermion fields. Up to
this point, there is no difference compared with general relativity at
all. Even non-relativistic Newtonian mechanics can be written down in
a coordinate-free fashion and shares all the properties mentioned
here, see, for example p.~300 of~\cite{MiTh73}.

In the non-generally relativistic treatment of field theories, for
example of Yang--Mills theory, the space-time metric $g$ is, however,
non-dynamical. This allows us to fix a special pair $(M,g)$ forever
and to study the isometries of the Riemannian manifold $(M,g)$, \ie\
those diffeomorphisms $f\colon M\to M$ for which $f^\ast g=g$. In the
common non-generally relativistic terminology, these isometries are
often called \emph{active transformations} because they `actively'
move the fields $A$ and $\psi$ relative to the fixed background
$(M,g)$. These active transformations do not form a gauge symmetry
because one can actually measure whether some object in the laboratory
has been translated or not. Of course, the laboratory is here viewed
as fixed with respect to the background $(M,g)$. If $(M,g)$ does not
have enough isometries, one may resort to the study of (local) Killing
vector fields, but the interpretation would not change.

The notion of active transformation requires an \emph{a priori}
decomposition of the physical fields into those that form the
\emph{background}, usually the space-time metric, which is
non-dynamical, plus other fields that live on this background and
which are treated as dynamical. The term \emph{passive transformation}
usually refers to the coordinate changes via transition functions, for
example to the transition from Cartesian to polar coordinates in some
flat geometry. The space-time diffeomorphisms $f\colon M\to M^\prime$
that relate two representatives $(M,g,A,\psi)$ and
$(M^\prime,g^\prime,A^\prime,\psi^\prime)$ of the same physical
configuration and which always form a gauge symmetry of the theory, do
not have any special name in the jargon of non-generally relativistic
physics.

The standard treatment of non-generally relativistic field theories
completely focuses on the active transformations and does not discuss
the generic diffeomorphism gauge symmetry which is nevertheless
present. In full general relativity, however, the metric is considered
dynamical so that the separation into a Riemannian manifold plus
fields that live is this geometry, is no longer available. The active
transformations of non-generally relativistic physics therefore have
no correspondence in the full theory of general relativity.

% ==============================================================================
%
\section{Path integrals}
%
% ==============================================================================
\label{sect_pathint}

In this section, we review the formal properties of path integrals for
general relativity (Subsection~\ref{sect_formal}) and exhibit their
close relationship with the axioms of TQFT, more precisely
$C^\infty$-QFT~\cite{At88} (Subsection~\ref{sect_tqft}). The
connection of $C^\infty$-QFT with theories that have diffeomorphisms
as a gauge symmetry, is already familiar from Witten's work on
Chern--Simons theory and knot invariants~\cite{Wi89} although in three
dimensions, the equivalence classes of smooth manifolds up to
diffeomorphism are in one-to-one correspondence with those of
topological manifolds up to homeomorphism. Barrett, Crane and
Baez--Dolan~\cite{Ba95,Cr95,BaDo95} have explicitly proposed
$C^\infty$-QFT as a framework for the quantization of general
relativity in $d=3+1$. We here recall the central ideas of this
connection and outline its relationship with the classification of
manifolds (Subsection~\ref{sect_invariants}). In the subsequent
sections, we explain why this framework is highly dimension-dependent
and why the diffeomorphism gauge symmetry alone is already sufficient
to single out $d=3+1$. We finally speculate about an extension of the
framework of $C^\infty$-QFT in order to better deal with the notion of
time in general relativity (Subsection~\ref{sect_corners}).

%------------------------------------------------------------------------------
\subsection{Formal properties}
%------------------------------------------------------------------------------
\label{sect_formal}

Path integral quantizations of general relativity, see, for
example~\cite{Mi57}, are supposed to share the following formal
properties. The subsequent discussion is purely heuristic.

\begin{figure}[t]
\begin{center}
\input{pstex/transition.pstex_t}
\end{center}
\mycaption{fig_transition}{%
  A $d$-manifold $M$ with boundary $\del M=\Sigma_1\dot\cup\Sigma_2$.}
\end{figure}

\paragraph{Hilbert spaces.}

We associate Hilbert spaces $\mathcal{H}(\Sigma)$ with closed
$(d-1)$-manifolds $\Sigma$. Such a $(d-1)$-manifold $\Sigma$
represents, for example, a space-like hyper-surface on which the
canonical variables of the theory are defined. Choosing a
polarization, we have a position representation
$\mathcal{H}=L^2(\mathcal{A})$ with suitable wave functionals on 
some set $\mathcal{A}$ of canonical coordinates. For definiteness, let
us imagine that we choose a \emph{connection representation} so that
$\mathcal{A}$ denotes the set of all connections $A|_\Sigma$.

\paragraph{Transition amplitudes.}

The path integral is supposed to describe transition amplitudes from
$\mathcal{H}(\Sigma_1)$ to $\mathcal{H}(\Sigma_2)$ for hyper-surfaces
$\Sigma_1$ and $\Sigma_2$, by summing over all histories of the
gravitational field, \ie\ over all space-time geometries, that
interpolate between the `initial' $\Sigma_1$ and the `final'
hyper-surface $\Sigma_2$. We therefore choose a $d$-manifold $M$ whose
boundary $\del M=\Sigma_1\dot\cup\Sigma_2$ is the disjoint union of
$\Sigma_1$ and $\Sigma_2$ in order to support these histories
(Figure~\ref{fig_transition}).

For connection eigenstates
$\ket{A|_{\Sigma_j}}\in\mathcal{H}(\Sigma_j)$, $j=1,2$, the transition
amplitudes are given by the (suitably normalized) path integral,
\begin{equation}
\label{eq_pathint}
  \bra{A|_{\Sigma_2}}T(M)\ket{A|_{\Sigma_1}} =
    \int_{A|_{\Sigma_1},A|_{\Sigma_2}}\mathcal{D}A\,\mathcal{D}e\,
    \exp\bigl(\frac{i}{\hbar}S[A,e]\bigr).
\end{equation}
These are the matrix elements of the linear \emph{transition map}
$T(M)\colon\mathcal{H}(\Sigma_1)\to\mathcal{H}(\Sigma_2)$. The path
integration $\int\mathcal{D}A\,\mathcal{D}e$ comprises the integration
over all connections $A$ compatible with the boundary conditions,
$A|_{\Sigma_1}$ and $A|_{\Sigma_2}$, and over all coframe fields $e$.

Notice that the connection representation for the Hilbert spaces
$\mathcal{H}(\Sigma)$ is compatible with our choice of boundary
conditions made above, namely to fix the connection $A|_{\del M}$ on
the boundary. We work with `real' time so that there is an $i$ in the
exponent. For generic states
$\ket{\psi(A|_{\Sigma_j})}\in\mathcal{H}(\Sigma_j)$, $j=1,2$, the path
integral reads,
\begin{equation}
\label{eq_matrixel}
  \bra{\psi(A|_{\Sigma_2})}T(M)\ket{\psi(A|_{\Sigma_1})} =
    \int\mathcal{D}A\,\mathcal{D}e\,
    \overline{\psi(A|_{\Sigma_2})}\psi(A|_{\Sigma_1})\,\exp\bigl(\frac{i}{\hbar}S[A,e]\bigr),
\end{equation}
where the integration over connections $A$ is now unrestricted.

\paragraph{Gauge symmetry.}

All manifolds and functions are assumed to be smooth, and everything
should be specified only `up to local Lorentz transformations' and `up
to diffeomorphisms' in a suitable way in order to implement the gauge
symmetry of general relativity. The Hilbert spaces therefore represent
states of spatial $(d-1)$-geometries up to diffeomorphism, \etc.

%------------------------------------------------------------------------------
\subsection{Axioms of $C^\infty$-QFT}
%------------------------------------------------------------------------------
\label{sect_tqft}

\begin{figure}[t]
\begin{center}
\input{pstex/composition.pstex_t}
\end{center}
\mycaption{fig_composition}{%
  Two $d$-manifolds $M_1$, $M_2$ with boundaries $\del
  M_1=\alignidx{\Sigma_1^\ast\dot\cup\Sigma_2}$ and $\del
  M_2=\alignidx{\Sigma_2^\ast\dot\cup\Sigma_3}$ glued together along
  their common boundary component $\Sigma_2$. The result is
  $M=M_1\cup_{\Sigma_2} M_2$.} 
\end{figure}

The framework sketched above is remarkably similar to the axiomatic
definition of $C^\infty$-QFT~\cite{At88}. The main advantage of
axiomatic $C^\infty$-QFT over Subsection~\ref{sect_formal}, and almost
the only difference, is that the diffeomorphism gauge symmetry is
carefully taken into account. We give here a slightly simplified
account of the original axioms. Note that the condition of
finite-dimensionality guarantees that everything is well defined, but
may finally have to be relaxed for general relativity. In the
following, all manifolds are smooth and oriented, and all
diffeomorphisms are orientation preserving. A $d$-dimensional
$C^\infty$-QFT,
\begin{myitemize}
\item[(S1)] 
  assigns to each closed $(d-1)$-manifold $\Sigma$ a
  [finite-dimensional] complex vector space $\mathcal{H}(\Sigma)$, and
\item[(S2)]
  assigns to each compact $d$-manifold $M$ with boundary $\del M$ a
  vector $T(M)\in\mathcal{H}(\del M)$,
\end{myitemize}
such that the following axioms (A1)--(A4) hold.
\begin{myitemize}
\item[(A1)] 
  The vector space of some opposite oriented manifold
  $\Sigma^\ast$ is the dual vector space,
  $\mathcal{H}(\Sigma^\ast)={\mathcal{H}(\Sigma)}^\ast$ (Bra-vectors
  live in the space dual to that of ket-vectors).
\item[(A2)]
  The vector space associated with a disjoint union of
  $(d-1)$-manifolds $\Sigma_1,\Sigma_2$ is the tensor product,
  $\mathcal{H}(\Sigma_1\dot\cup\Sigma_2)\cong\mathcal{H}(\Sigma_1)\otimes\mathcal{H}(\Sigma_2)$
  (The Hilbert space of a composite system is the tensor product of
  the spaces of the constituents).
\item[(A3)]
  For each diffeomorphism $f\colon\Sigma_1\to\Sigma_2$ between closed
  $(d-1)$-manifolds $\Sigma_j$, there exists a linear isomorphism
  $f^\prime\colon\mathcal{H}(\Sigma_1)\to\mathcal{H}(\Sigma_2)$. If
  $f$ and $g\colon\Sigma_2\to\Sigma_3$ are diffeomorphisms, then
  ${(g\circ f)}^\prime=g^\prime\circ f^\prime$. Furthermore, for each
  diffeomorphism $f\colon M_1\to M_2$ between compact $d$-manifolds
  $M_j$ with boundary $\del M_j=\Sigma_j$, it is required that
  $T(M_2)=f|_{\Sigma_1}^\prime\,(T(M_1))$, where
  $f|_{\Sigma_1}\colon\Sigma_1\to\Sigma_2$ denotes the restriction to
  the boundary (This is the implementation of the diffeomorphism gauge
  symmetry).
\end{myitemize}
The axioms~(A1) and~(A2) imply that any $d$-manifold $M$ whose
boundary is a disjoint union of the form $\del
M=\alignidx{\Sigma_1^\ast\dot\cup\Sigma_2}$, is assigned a vector
$T(M)\in\mathcal{H}(\Sigma_1)^\ast\otimes\mathcal{H}(\Sigma_2)\cong\Hom_\C(\mathcal{H}(\Sigma_1),\mathcal{H}(\Sigma_2))$,
\ie\ a linear map
$T(M)\colon\mathcal{H}(\Sigma_1)\to\mathcal{H}(\Sigma_2)$ (These are
the desired transition maps~\eqref{eq_matrixel}).
\begin{myitemize}
\item[(A4)] If compact $d$-manifolds $M_1$, $M_2$ with boundaries
  $\del M_1=\alignidx{\Sigma_1^\ast\dot\cup\Sigma_2}$ and $\del
  M_2=\alignidx{\Sigma_2^\ast\dot\cup\Sigma_3}$ are glued together
  along their common boundary component $\Sigma_2$
  (Figure~\ref{fig_composition}), the resulting manifold
  $M=M_2\cup_{\Sigma_2} M_1$ yields the composition of the transition
  maps, $T(M_2\cup_{\Sigma_2} M_1)=T(M_2)\circ T(M_1)$ which is a
  linear map from $\mathcal{H}(\Sigma_1)$ to $\mathcal{H}(\Sigma_3)$.
\end{myitemize}
Notice that the local Lorentz symmetry was not explicitly
mentioned. It is usually taken into account by an appropriate choice
of vector spaces $\mathcal{H}(\Sigma)$. The change of spatial topology
(more precisely of the spatial smooth manifold) is possible in this
framework if $\Sigma_1$ and $\Sigma_2$ in Figure~\ref{fig_transition}
are not diffeomorphic. The precise form of this topology change is
encoded in the choice of $M$. The $d$-manifold $M$ itself, however, is
always fixed. Generalizations that include the `superposition' of
different $d$-manifolds $M$ are not covered by the correspondence
principle and would therefore require new physical assumptions.

%------------------------------------------------------------------------------
\subsection{Invariants of smooth manifolds}
%------------------------------------------------------------------------------
\label{sect_invariants}

\begin{figure}[t]
\begin{center}
\input{pstex/projector.pstex_t}
\end{center}
\mycaption{fig_projector}{% 
  Gluing together two cylinders $\Sigma\times I$ along one of their
  boundaries $\Sigma$ gives another cylinder (up to diffeomorphism).}
\end{figure}

The axioms have some interesting consequences.
\begin{myitemize}
\item[1.]
  Axiom~(A4) applied to the case of disjoint manifolds $M_1$ and
  $M_2$, \ie\ $\del M_1=\Sigma_1$, $\del M_2=\Sigma_2$,
  $\Sigma_1\cap\Sigma_2=\emptyset$, implies that the associated vector
  is just the tensor product of vectors, $T(M_1\dot\cup M_2)=T(M_1)\otimes
  T(M_2)\in\mathcal{H}(\Sigma_1)\otimes\mathcal{H}(\Sigma_2)$ (State
  of a system that is composed from independent constituents).
\item[2.]
  Axiom~(A2) applied to $\Sigma_2=\emptyset$ shows that
  $\mathcal{H}(\emptyset)=\C$ or otherwise all $\mathcal{H}(\Sigma)$
  are null.
\item[3.]
  Axiom~(A4) applied to $M_1=M_2=\emptyset$ implies that 
  $T(\emptyset)\in\mathcal{H}(\emptyset)=\C$ satisfies
  $T(\emptyset)=1$ or otherwise all $T(M)=0$.
\item[4.]
  Axiom~(A4) applied to a cylinder $M=\Sigma\times I$ where $\Sigma$
  is a closed $(d-1)$-manifold and $I=[0,1]$ the unit interval
  (Figure~\ref{fig_projector}), shows that the transition map
  $T(M)\colon\mathcal{H}(\Sigma)\to\mathcal{H}(\Sigma)$ is a
  projection operator, $T(M)^2=T(M)$. For any $d$-manifold that can be
  written as a cylinder glued to something else, this projector is
  effective so that what matters is only its image. Therefore, one
  often imposes the additional condition that $T(\Sigma\times
  I)=\id_{\mathcal{H}(\Sigma)}$.   
\end{myitemize}

\begin{figure}[t]
\begin{center}
\input{pstex/partition.pstex_t}
\end{center}
\mycaption{fig_partition}{% 
  If a compact $d$-manifold $M$ has got some boundary $\del
  M=\Sigma_1^\ast\dot\cup\Sigma_2$ whose components are diffeomorphic,
  one can use the diffeomorphism $f\colon\Sigma_2\to\Sigma_1$ in order
  to glue them together and form the closed manifold $M_f$.}
\end{figure}

It is well-known~\cite{At88} that the appearance of these projection
operators is closely related to the vanishing of the Hamiltonian,
$H=0$, in the corresponding canonical formulation of the theory. This
is in fact a prominent feature of general relativity~\cite{Mi57} as
soon as the $3+1$ splitting for the Hamiltonian formulation is done
with respect to coordinate time. In this sense, both axiomatic
$C^\infty$-QFT and canonical general relativity are `non-dynamical'
theories. This is in both cases a direct consequence of the
diffeomorphism gauge symmetry.

\begin{myitemize}
\item[5.]
  For closed $d$-manifolds, $\del M=\emptyset$, so that
  $T(M)\in\mathcal{H}(\emptyset)=\C$. Therefore, $T(M)$ is just a number
  that depends on the diffeomorphism class of the smooth manifold
  $M$. It is an invariant of smooth manifolds! In analogy with
  statistical mechanics, this number is called the
  \emph{partition function} of $M$ and denoted by $Z(M):=T(M)$.
\item[6.]
  More generally, for a compact $d$-manifold $M$ whose boundary is of the
  form $\del M=\alignidx{\Sigma_1^\ast\dot\cup\Sigma_2}$ where
  $\Sigma_1$ and $\Sigma_2$ are diffeomorphic, one can use such a
  diffeomorphism $f\colon\Sigma_2\to\Sigma_1$ in order to glue $M$ to
  itself at its boundary so that one obtains a closed manifold $M_f$
  (Figure~\ref{fig_partition}). In this case, by axiom~(A3), there is
  an isomorphism of vector spaces,
  $f^\prime\colon\mathcal{H}(\Sigma_2)\to\mathcal{H}(\Sigma_1)$ so
  that the partition function of $M_f$ can be written as the trace,
\begin{equation}
  Z(M_f) = \tr_{\mathcal{H}(\Sigma_1)}(f^\prime\circ T(M)).
\end{equation}
  In the language of the path integral~\eqref{eq_pathint}, this
  calculation of the trace corresponds to integrating over all
  possible boundary conditions on the identified
  $\Sigma_1\equiv_f\Sigma_2$. If actually $\Sigma_1=\Sigma_2=\Sigma$
  and $f=\id_{\mathcal{H}(\Sigma)}$, 
\begin{equation}
\label{eq_partition}
  Z(M_{\id_\Sigma}) = \tr_{\mathcal{H}(\Sigma)} T(M)
  = \int\mathcal{D}A\,\mathcal{D}e\,
    \exp\bigl(\frac{i}{\hbar}S[A,e]\bigr).
\end{equation}
  This formula for $Z(M_{\id_\Sigma})$ in terms of the unrestricted
  path integral (over the manifold $M_{\id_\Sigma}$) motivates the term
  partition function.
\end{myitemize}

The most important implications are the items~(5.) and~(6.). It is
essentially a consequence of the diffeomorphism gauge symmetry that
the partition function $Z(M)$ which can be computed from the path
integral, is an invariant of smooth manifolds. Although the partition
function itself does not have any physical meaning, it is therefore
mathematically very valuable. The physically relevant objects are the
matrix elements~\eqref{eq_matrixel}. They are more general than just
the partition function which is their trace, but for the beginning,
let us focus on the partition function. In the subsequent sections, we
ask in which space-time dimensions we can expect interesting theories,
for example,
\begin{myitemize}
\item
  Are smooth manifolds up to diffeomorphism already characterized by
  simpler structures, for example, by their underlying topological
  manifolds up to homeomorphism or even up to homotopy equivalence?
\item
  In which space-time dimensions is it possible to compute the
  partition function more efficiently in a purely combinatorial
  context using suitable triangulations of the space-time manifold?
  Which role do these triangulations play in general relativity?
\end{myitemize}

%------------------------------------------------------------------------------
\subsection{Extensions of the framework of $C^\infty$-QFT}
%------------------------------------------------------------------------------
\label{sect_corners}

\paragraph{Dimensionality.}

As mentioned before, for general relativity it may be necessary to
drop the words `finite-dimensional' from~(S1) above. In this case, one
has to be more careful with the notion of dual vector space and with
the construction of the traces which may become infinite. We
illustrate this below in Section~\ref{sect_three} for the example
of quantum gravity in $d=2+1$.

\paragraph{Hermitean structures.}

A Hermitean scalar product on the vector spaces $\mathcal{H}(\Sigma)$
gives rise to an isomorphism
$\mathcal{H}(\Sigma^\ast)={\mathcal{H}(\Sigma)}^\ast\cong\overline{\mathcal{H}(\Sigma)}$
where $\overline{\mathcal{H}(\Sigma)}$ denotes the vector space
$\mathcal{H}(\Sigma)$ with the complex conjugate action of the
scalars. In this case, it is possible to add the axiom,
\begin{myitemize}
\item[(A5)]
  For each $d$-manifold $M$, the vector $T(M)\in\mathcal{H}(\del M)$ 
  satisfies $T(M^\ast)=\overline{T(M)}$.
\end{myitemize}
If $\del M=\alignidx{\Sigma_1^\ast\dot\cup\Sigma_2}$ and
$T(M)\colon\mathcal{H}(\Sigma_1)\to\mathcal{H}(\Sigma_2)$ is viewed as
a linear map, this axiom relates orientation reversal of space-time
with the hermitean adjoint of the transition map,
$T(M^\ast)={T(M)}^\dagger$.

\paragraph{Unitarity.}

Assume that the axiom~(A5) is satisfied. If the transition map
$T(M)\colon\mathcal{H}(\Sigma_1)\to\mathcal{H}(\Sigma_2)$ for $\del
M=\alignidx{\Sigma_1^\ast\dot\cup\Sigma_2}$ was unitary, \ie\
${T(M)}^\dagger={T(M)}^{-1}$, orientation reversal of $M$ would just
invert the transition map, $T(M^\ast)={T(M)}^{-1}$. It is tempting to
think that orientation reversal was in this way related with time
reversal.

Unitarity in quantum theory, however, expresses the conservation of
probability with respect to a global time parameter and therefore
cannot be expected to hold in general relativity without additional
assumptions.

Consider, for example, the path integral~\eqref{eq_pathint} for the
$d$-manifold $M$ of Figure~\ref{fig_transition} where the boundary
data $A|_{\Sigma_1}$ and $A|_{\Sigma_2}$ impose space-like geometries
on $\Sigma_1$ and $\Sigma_2$, respectively. The path integral will
generically contain histories in which $\Sigma_2$ is in the causal
future of $\Sigma_1$ and also some in which $\Sigma_2$ is in the past
of $\Sigma_1$, unless we impose additional conditions.

Following the ideas of Oeckl~\cite{Oe03}, one can consider more
general manifolds of the form $M=S\times I$ where $S$ is a compact
$(d-1)$-manifold \emph{with} boundary $\del S\neq\emptyset$. The
boundary $\del M$ then consists of $\del M=(S\times\{0\})\cup
(S\times\{1\}) \cup(\del S\times I)$. The idea is now to impose
space-like geometries on $S\times\{0\}$ (initial preparation of the
experiment) and $S\times\{1\}$ (final measurement of the outcome) and
in addition a time-like geometry on $\del S\times I$ which represents
the clock in the classical laboratory that surrounds the quantum
experiment and which ensures that $S\times\{1\}$ is in the future of
$S\times\{0\}$.

Notice that this is the situation in which textbook quantum theory
makes sense without conceptual extensions and in which it has been
confirmed experimentally: the quantum experiment is limited both in
size and duration, and all measurements are performed by classical
observers who use classical clocks.

\paragraph{Corners and higher level.}

If we just plugged this choice of $M=S\times I$ into the axioms of
Section~\ref{sect_tqft}, we would get a single Hilbert space
associated with the boundary $\del M$. This is not quite what we
want. We would rather prefer to obtain Hilbert spaces for the boundary
components $S\times\{0\}$ and $S\times\{1\}$ between which the
transition map acts, but not for $\del S\times I$. The framework of
higher level TQFT or TQFT with corners, see for example~\cite{BaDo95},
might be appropriate to treat this situation. We do not go into
details here, but rather focus on the structure of manifolds in the
subsequent sections.

% ==============================================================================
%
\section{Classification of manifolds}
%
% ==============================================================================
\label{sect_manifolds}

Given some path integral~\eqref{eq_pathint} of general relativity in
$d$-dimensional space-time, let us consider the partition function
$Z(M)$ of~\eqref{eq_partition}, assuming for a moment that it is well
defined and can be computed for some class of space-time manifolds.

Whenever $Z(M)\neq Z(M^\prime)$, then $M$ and $M^\prime$ are not
diffeomorphic. The partition function therefore forms a tool for the
classification of smooth manifolds up to diffeomorphism. Although the
framework of general relativity is by definition that of smooth
manifolds and smooth maps (the Einstein equations are differential
equations after all), it is instructive to contrast it with other
types of manifolds. Barrett~\cite{Ba95} has already remarked that the
space-time dimension $d=3+1$ plays a special role. Let us explain in
more detail why. The presentation in this section is a rather informal
overview. The detailed definitions and theorems to which we refer, can
be found in the Appendix.

\paragraph{Topological manifolds up to homeomorphism.}

Each smooth manifold has got an underlying topological manifold
(Section~\ref{sect_manifold}), and any two diffeomorphic smooth
manifolds have homeomorphic underlying topological manifolds. Can we
use the information about topological manifolds up to homeomorphism in
order to classify smooth manifolds up to diffeomorphism?

The answer is `yes' in space-time dimension $d\leq 2+1$, but `no' in
general. Indeed, in $d\leq 2+1$, each topological manifold admits a
differentiable structure, and any two homeomorphic topological
manifolds have differentiable structures so that the resulting smooth
manifolds are diffeomorphic.

This result is one of the explanations of why quantum general
relativity in $d=2+1$ is particularly simple. In fact, its path
integral quantization is closely related to a $C^0$-QFT, using
topological manifolds up to homeomorphism, and exploiting the fact
that in $d=2+1$, there is no difference between $C^0$- and
$C^\infty$-QFTs. Examples of such theories are given in
Section~\ref{sect_three} below, but before we can state them, we need
some more theoretical background.

In $d\geq 3+1$, no analogous result is available. There exist
countably infinite families of (compact) smooth
$4$-manifolds~\cite{FrMo88,OkVe86} which are pairwise
non-diffeomorphic, but which have homeomorphic underlying topological
manifolds. There is therefore a considerable discrepancy between
$C^\infty$- and $C^0$-QFTs in $d=3+1$ space-time dimensions.

The most striking result even concerns the standard space
$\R^4$~\cite{Ta87,GoSt99}.

\begin{theorem}
Consider the topological manifold $\R^d$, $d\in\N$.
\begin{myitemize}
\item
  If $d\neq 4$, then there exists a differentiable structure for
  $\R^d$ which is unique up to diffeomorphism.
\item
  If $d=4$, then there exists an uncountable family of pairwise
  non-diffeomorphic differentiable structure for $\R^d$.
\end{myitemize}
\end{theorem}

The differentiable structure of $\R^4$ induced from
$\R\times\R\times\R\times\R$ is called \emph{standard} and the others
\emph{exotic}.

Non-uniqueness of differentiable structures persists in higher
dimensions, for example, there are $28$ inequivalent differentiable
structures on the topological sphere $S^7$, or $992$ inequivalent
differentiable structures on $S^{11}$~\cite{KeMi63}, but in dimension
$d\geq 4+1$ ($d\geq 5+1$ if the manifold has a non-empty boundary),
there never exists more than a finite number of non-diffeomorphic
differentiable structures on the same underlying topological
manifold. The space-time dimension $d=3+1$ is distinguished by the
feature that there can exist an infinite number of homeomorphic, but
pairwise non-diffeomorphic compact smooth manifolds.

\paragraph{Topological manifolds up to homotopy equivalence.}

There is another way of classifying topological manifolds. This
relation is known as \emph{homotopy equivalence}. Two topological
manifolds $M$, $N$ are called homotopy equivalent if there exists a
pair of continuous maps $f\colon M\to N$ and $g\colon N\to M$ such
that $f\circ g$ is \emph{homotopic} to the identity map $\id_M$ of
$M$, \ie\ it `can be continuously deformed' to $\id_M$, and $g\circ f$
is homotopic to $\id_N$ (see Appendix~\ref{app_homotopytype} for
details).

The concept of homotopy equivalence of topological manifolds is weaker
than that of homeomorphism: any two homeomorphic topological manifolds
are also homotopy equivalent.

The converse implication is true, for example, in $d\leq 1+1$, \ie\
any two compact topological manifolds that are homotopy equivalent,
are also homeomorphic. But it does not hold in $d=2+1$: there exist
compact topological $3$-manifolds which are homotopy equivalent, but
not homeomorphic.

In $d\leq 1+1$, quantum general relativity is therefore even simpler
than in $d=2+1$. It is not only given by a $C^0$-QFT (as opposed to a
generic $C^\infty$-QFT), but even by what one could call an hQFT (TQFT
up to homotopy equivalence\footnote{The name \emph{homotopy quantum
field theory} and the abbreviation HQFT are already gone for a
different concept.}). Quantum general relativity in $d=2+1$, in
contrast, is not an hQFT. A topological invariant closely related to
its partition function has been confirmed to distinguish homotopy
equivalent topological $3$-manifolds that are not
homeomorphic. Quantum general relativity in $d=2+1$ is thus as generic
as topology allows.

\paragraph{Piecewise-linear manifolds up to PL-isomorphism.}

\begin{figure}[t]
\begin{center}
\input{pstex/simplex.pstex_t}
\end{center}
\mycaption{fig_simplex}{% 
  $k$-simplices in $\R^3$.}
\end{figure}

Before we can cite the next relevant classification result, we have to
introduce yet another type of manifolds: piecewise-linear (PL-)
manifolds (see Appendix~\ref{app_pl} for details).

A \emph{$k$-simplex} in standard space $\R^d$ is the smallest convex
set that contains $k+1$ points that span a $k$-dimensional hyperplane
(Figure~\ref{fig_simplex}). A \emph{polyhedron} is a locally finite
union of simplices. $\R^d$ itself, for example, is a polyhedron. A
\emph{piecewise-linear} map is a map $f\colon P\to Q$ between
polyhedra which maps simplices onto simplices. We can now obtain
another type of manifold by restricting the transition functions of a
topological manifold to piecewise-linear functions.

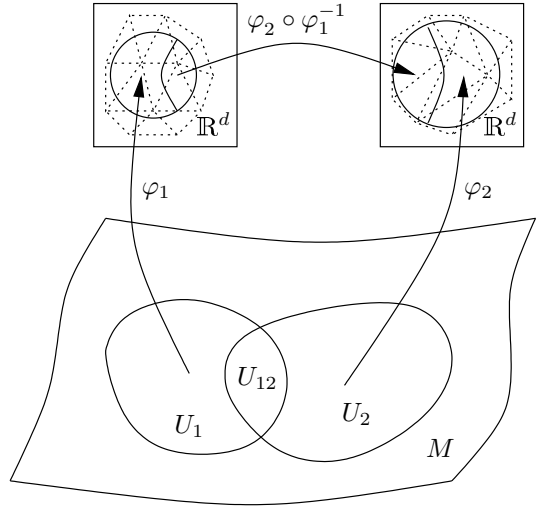
A \emph{piecewise-linear} (PL-) manifold is a topological manifold
such that all transition functions are piecewise-linear. This is
illustrated in Figure~\ref{fig_plmanifold}. Notice that not $M$ itself
is triangulated, but rather the coordinate systems are. PL-manifolds
are classified up to PL-isomorphism (Appendix~\ref{app_pl}), \ie\ up
to homeomorphisms that are piecewise-linear in coordinates.

\begin{figure}[t]
\begin{center}
\input{pstex/plmanifold.pstex_t}
\end{center}
\mycaption{fig_plmanifold}{% 
  A piecewise-linear (PL-) manifold.}
\end{figure}

\paragraph{Triangulations of smooth manifolds.}

If some topological manifold admits both a piece\-wise-linear and a
smooth structure, satisfying a compatibility condition (see
Appendix~\ref{app_smoothing} for details), we say that the
differentiable structure is a \emph{smoothing} of the piecewise-linear
structure.

There is a close relationship between smooth and piecewise-linear
manifolds given by Whitehead's theorem: For each smooth manifold $M$,
there exists a PL-manifold $M_\mathrm{PL}$, called its \emph{Whitehead
triangulation}, so that $M$ is diffeomorphic to a smoothing of
$M_\mathrm{PL}$. $M_\mathrm{PL}$ is unique up to PL-isomorphism.

Whitehead's theorem is therefore a license to triangulate
space-time. But does the Whitehead triangulation capture all features
of the given smooth manifold? The answer is `yes', at least in
space-time dimension $d\leq 5+1$: each PL-manifold admits a smoothing,
and the resulting smooth manifold is unique up to diffeomorphism. The
equivalence classes of smooth manifolds up to diffeomorphism are
therefore in one-to-one correspondence with those of PL-manifolds up
to PL-isomorphism. We are free to choose either framework at any time.

We therefore know that the path integral of general relativity in
$d\leq 5+1$ is closely related to a TQFT that is defined for
PL-manifolds up to PL-isomorphism, \ie\ to a PL-QFT. Whitehead
triangulations provide us with a way of `discretizing' space-time
which is not merely some approximation nor introduces a physical
cut-off, but which is rather exact up to diffeomorphism. 

\paragraph{Combinatorial manifolds}

\begin{figure}[t]
\begin{center}
\input{pstex/combinatorial.pstex_t}
\end{center}
\mycaption{fig_combmanifold}{% 
  Each piecewise-linear (PL-) manifold $M$ is PL-isomorphic to some
  combinatorial manifold $P$.}
\end{figure}

\begin{figure}[t]
\begin{center}
\input{pstex/pachner.pstex_t}
\end{center}
\mycaption{fig_pachner}{% 
  Pachner moves in $d=2$: (a) The $1\leftrightarrow 3$ move subdivides
  a triangle into three. (b) the $2\leftrightarrow 2$ move joins two
  triangles and then splits the result in a different way. Pachner
  moves in $d=3$: (c) the $1\leftrightarrow 4$ move subdivides
  one tetrahedron into four. (d) the $2\leftrightarrow 3$ move changes
  the subdivision of a diamond from two tetrahedra to three ones
  (glued along the dotted line in the bottom picture). 
}
\end{figure}

Whitehead triangulations are widely used in topology because they
facilitate efficient computations which are most conveniently
performed in a purely combinatorial language.

It is known that each $d$-dimensional PL-manifold $M$ is PL-isomorphic
to a single polyhedron $P$ in $\R^n$ for some $n$
(Figure~\ref{fig_combmanifold}). Such a polyhedron which itself forms
a PL-manifold, is called a \emph{combinatorial manifold} (see
Appendix~\ref{app_combinatorial} for details) or a \emph{global
triangulation} of $M$. If $M$ is compact, $P$ can be described in
terms of a finite number of simplices. In order to characterize $M$ up
to PL-isomorphism, it suffices to characterize the PL-manifold $P$ up
to PL-isomorphism.

A very convenient way of stating the condition of PL-isomorphism, and
for our purposes the best intuition, is provided by Pachner's
theorem. We give here the version for closed manifolds: any two closed
combinatorial manifolds are PL-isomorphic if and only if they are
related by a finite sequence of Pachner moves.

Pachner moves are local modifications of the triangulation by joining
simplices or by splitting some polyhedra up into smaller
pieces. Figure~\ref{fig_pachner} shows the Pachner moves for closed
manifolds in $d=1+1$ and $d=2+1$. A systematic way of listing the
moves in any dimension is explained in
Appendix~\ref{app_abstract}. Pictures for $d=3+1$ can be found
in~\cite{CaKa99,Ma99}, and the moves for manifolds with boundary
in~\cite{Pa91}.

The following procedure is now available in order to decide whether
two given smooth $d$-manifolds, $d\leq 5+1$, are diffeomorphic. Start
with two closed smooth manifolds $M^{(1)}$ and $M^{(2)}$. Construct
their Whitehead triangulations $M^{(1)}_\mathrm{PL}$ and
$M^{(2)}_\mathrm{PL}$. Find combinatorial manifolds $P^{(1)}$ and
$P^{(2)}$ which are PL-isomorphic to $M^{(1)}_\mathrm{PL}$ and
$M^{(2)}_\mathrm{PL}$, respectively. $M^{(1)}$ and $M^{(2)}$ are
diffeomorphic if and only if $P^{(1)}$ and $P^{(2)}$ are related by a
finite sequence of Pachner moves.

\paragraph{Scenario for quantum gravity.}

We have reached a first goal: the diffeomorphism gauge symmetry of
general relativity on a closed space-time manifold has been translated
into a purely combinatorial problem involving triangulations that
consist of only a finite number of simplices, and their manipulation
by finite sequences of Pachner moves. If not only the partition
function, but also the full path integral of general relativity in
$d\leq 5+1$ is given by a PL-QFT, we know that all observables are
invariant under Pachner moves.

\begin{table}
\begin{center}
\begin{tabular}{|c|c|c|c|}
\hline
$d\leq 2$  & $d=3$      & $d=4$      & $d\geq 5$  \\
\hline
\hline
$C^\infty$ & $C^\infty$ & $C^\infty$ & $C^\infty$ \\[-1mm]
           &            &            & \dotfill   \\
PL         & PL         & PL         & PL         \\[-1mm]
           &            & \hrulefill & \dotfill   \\
Top        & Top        & Top        & Top        \\[-1mm]
           & \hrulefill &            & \dotfill   \\
htpy       & htpy       & htpy       & htpy       \\
\hline
\end{tabular}
\end{center}
\caption{
\label{tab_classification}
The relationship between the classifications of various types of
manifolds of dimension $d$: smooth manifolds up to diffeomorphism
($C^\infty$), piecewise-linear manifolds up to PL-isomorphism (PL),
topological manifolds up to homeomorphism (Top), and topological
manifolds up to homotopy equivalence (htpy). Equivalence of manifolds
of the type shown in one row implies equivalence of manifolds of the
type shown in the rows below. If the rows are not separated by any
line, the converse implication holds as well. If the rows are
separated by a dotted line, the converse implication holds up to some
obstruction and ambiguity. A solid line indicates that the converse
implication is seriously violated. For details, we refer to the
Appendix, in particular to Appendix~\ref{app_comparison}. This table
is rather sketchy, and a number of subtleties have been suppressed, so
we ask the reader not to consider this table as a \emph{theorem}
without actually having read the small-print in the Appendix.  }
\end{table}

The partition function of quantum general relativity is an invariant
of PL-manifolds, too, and can be computed by purely combinatorial
methods for any given combinatorial manifold. A generic expression of
such a partition function is the \emph{state sum},
\begin{equation}
\label{eq_combinatorial}
  Z = \sum_{\{\,\mathrm{colourings}\,\}}\prod_{\{\,\rm{simplices}\,\}}(\mathrm{amplitudes}),
\end{equation}
where the sum is over all labellings of the simplices with elements of
some \emph{set of colours}, and the integrand is a number that can be
computed for each such labelling. In Section~\ref{sect_three} below, we
give examples and illustrate that the partition function of quantum
general relativity in $d=2+1$ is precisely of this form.

If quantum general relativity in $d=3+1$ is indeed a PL-QFT, the
following two statements which sound philosophically completely
contrary,
\begin{myitemize}
\item
  Nature is fundamentally smooth.
\item
  Nature is fundamentally discrete.
\end{myitemize}
are just two different points of view on the same underlying
mathematical structure: equivalence classes of smooth manifolds up to
diffeomorphism.

Further classification results which are compiled in the Appendix,
indicate that the partition function in $d\geq 4+1$ would be much less
interesting than in $d=3+1$ because smooth manifolds up to
diffeomorphism are already essentially classified by their underlying
topological manifold up to homeomorphism, subject to only finite
ambiguities in choosing a differentiable structure. In fact, one could
even use the homotopy type of space-time plus some additional
information about the structure of the tangent bundle. Quantum general
relativity in $d\geq 4+1$ would therefore be closely related to an
hQFT supplemented by additional data in order to specify the tangent
bundle and to resolve the ambiguities.

On the mathematical side, the path integral quantization of general
relativity is closely related to the problem of classifying smooth
manifolds up to diffeomorphism by classifying their Whitehead
triangulations up to PL-isomorphism. Precisely in $d=3+1$, topology is
rich enough to (potentially) provide infinitely many non-trivial
partition functions that are able to distinguish non-PL-isomorphic
PL-structures on the same underlying topological manifold. It is an
open problem in topology to construct these invariants.

Unless $d=2+1$ or $d=3+1$, the problem of constructing interesting
partition functions is finally dominated by the study of topological
manifolds up homotopy equivalence which would render general
relativity `suspiciously simple'. All these considerations apply to
the partition function~\eqref{eq_partition}, but not necessarily to
the matrix elements~\eqref{eq_matrixel}, \ie\ before tracing out the
boundary conditions. In the next section, we illustrate how the
examples in $d=2+1$ show that the expression for the state
sum~\eqref{eq_combinatorial} already suggest the appropriate Hilbert
spaces and boundary conditions. Finding the partition function is
therefore a key step.

Notice that we are not claiming that the universe is a smooth
$4$-manifold $M$ that has an `exotic', \ie\ non-standard,
differentiable structure. This may or may not be true, and the answer
to this question is independent of the considerations presented so
far. Some consequences of exotic differentiable structures on
space-time have been explored in~\cite{Br94,AsBr02}. The crucial
observation is rather that, just because the Einstein equations are
differential equations, the path integral of general relativity ought
to be sensitive to the differentiable structure of $M$, even if it is
merely the standard differentiable structure. It may eventually turn
out that in many cases the smooth structure on the boundary $\del M$
already determines the smooth structure of the entire $M$, \ie\ that
the smooth background can be viewed as part of the classical boundary
data.

Table~\ref{tab_classification} finally summarizes the classification
of smooth, PL- and topological manifolds in the various
dimensions. Refer to the Appendix for details.

\paragraph{Renormalization group transformations.}

The $1\leftrightarrow d+1$ Pachner move (Figure~\ref{fig_pachner})
always subdivides one $d$-simplex into $d+1$ $d$-simplices. It
obviously resembles the block spin transformations familiar from the
Statistical Mechanics treatment of renormalization, but is here
applied to the Whitehead triangulations of smooth manifolds as opposed
to discretizations of Riemannian manifolds. In our case, the simplices
do not have any metric size as there is no background geometry
associated with the smooth space-time manifold $M$.

In non-generally relativistic field theories on some given Riemannian
manifold $(M,g)$, renormalization is the comparison of the dynamical
scale of the theory, \ie\ the relevant correlation lengths, for any
given cut-off and bare parameters, with the scale of the background
metric $g$, in order to determine the relation between cut-off and
bare parameters for which the physical predictions are constant.

In general relativity, there is no background metric and therefore no
way of (and no need to) introduce a cut-off. The diffeomorphism gauge
symmetry implies the invariance of all observables under Pachner moves
so that one can say that the theory is readily renormalized or,
depending on the personal taste, that there is no need to renormalize
theories in which the geometry is dynamical. We stress that Whitehead
triangulations neither introduce a cut-off nor break any of the
symmetries. Refer to~\cite{Pf03} for the implications on the notion of
locality, on the appearance of the Planck scale and on the
compatibility with ideas in the context of the holographic principle.

% ==============================================================================
%
\section{Examples}
%
% ==============================================================================
\label{sect_three}

This section is a brief overview over some results on quantum general
relativity in $d=2+1$ space-time dimensions. For more details, see the
review articles~~\cite{Ba99,Or01,Pe03}. 

\paragraph{Turaev--Viro invariant.}

We have stressed above the importance of the partition function and
that it forms an invariant of smooth manifolds up to
diffeomorphism. In $d=2+1$, we know that we can equivalently study an
invariant of PL-manifolds up to PL-isomorphism which is given by a
state sum~\eqref{eq_combinatorial} for a closed combinatorial
manifold. The Turaev--Viro invariant~\cite{TuVi92} is such a state
sum,
\begin{equation}
\label{eq_tuvi}
  Z = \sum_{j\colon\Delta_1\to\{0,\frac{1}{2},1,\ldots,\frac{k-2}{2}\}}
  \Bigl(\prod_{\sigma_1\in\Delta_1}\dim_q j(\sigma_1)\Bigr)
  \Bigl(\prod_{\sigma_3\in\Delta_3}{\{6j\}}_q(\sigma_3)\Bigr).
\end{equation}
Here $k=1,\frac{3}{2},2,\ldots$ is a fixed half-integer, $\Delta_1$
and $\Delta_3$ denote the sets of $1$-simplices and $3$-simplices,
respectively. The sum is over all ways of colouring the $1$-simplices
$\sigma_1\in\Delta_1$ with half-integers
$j(\sigma_1)\in\{0,\frac{1}{2},1,\ldots,\frac{k-2}{2}\}$. The
\emph{quantum dimension} $\dim_q j(\sigma_1)$ can be computed for each
$j(\sigma_1)$, and the \emph{quantum-$6j$-symbol}
${\{6j\}}_q(\sigma_3)$ depends on the labels $j(\sigma_1)$ associated
with all $1$-simplices $\sigma_1$ in the boundary of each $3$-simplex
$\sigma_3$, see~\cite{TuVi92} for details. The half-integers
$j(\sigma_1)$ in fact characterize the finite-dimensional irreducible
representations of the quantum group $U_q(\mathfrak{sl}_2)$ for the
root of unity $q=e^{i\pi/k}$.

\paragraph{Ponzano--Regge model.}

The first connection with quantum gravity in $d=2+1$ can be seen in
the limit $k\to\infty$ in which $q\to 1$ and the quantum group
$U_q(\mathfrak{sl}_2)$ is replaced by the envelope
$U(\mathfrak{sl}_2)$ which is dual to the algebra of functions on the
local Lorentz group $\SU(2)=\Spin(3)$ (up to complexification). The
$j(\sigma_1)$ then characterize the finite-dimensional irreducible
representations of $\SU(2)$. In this limit, $Z$ agrees with the
partition function of the Ponzano--Regge model~\cite{PoRe68}, a
non-perturbative quantization of the toy model of general relativity
in $d=2+1$ with Riemannian signature $\eta=\diag(1,1,1)$. In the
limit, the partition function diverges and is no longer a
mathematically well-defined invariant, but the Pachner move invariance
of~\eqref{eq_tuvi} persists formally if one accepts to divide out
infinite factors.

For the model with the realistic Lorentzian signature $\eta=(-1,1,1)$,
$\Spin(3)$ is replaced by $\Spin(1,2)$ so that the representation
labels become continuous~\cite{Da01,Fr01}.

\paragraph{Cosmological constant.}

In the Ponzano--Regge model, the large spin limit of the $6j$-symbols
yields the connection with the classical action of general
relativity~\cite{PoRe68} and shows that the labels $j(\sigma_1)$
represent dynamically assigned lengths\footnote{In $d=2+1$, $\hbar G$
is the Planck length.} $\hbar G(j(\sigma_1)+\frac{1}{2})$ for the
$1$-simplices $\sigma_1$. The analogous argument for the Turaev--Viro
invariant indicates~\cite{MiTa92,Ba03} that the Turaev--Viro invariant
is the partition function of general relativity with Riemannian
signature and quantized positive cosmological constant
$\Lambda=4\pi^2/{(\hbar Gk)}^2$. The limit $k\to\infty$ then sends
$\Lambda\to0$ as expected.

\paragraph{TQFT.}

Although we have so far concentrated on the partition function, one
can easily read off from the state sum~\eqref{eq_tuvi} a consistent
choice of boundary fields and Hilbert spaces~\cite{TuVi92}: fix the
$j(\sigma_1)$ for all $1$-simplices in the boundary $\Sigma=\del M$ in
order to characterize a state. The Hilbert space $\mathcal{H}(\Sigma)$
then has a basis whose vectors are labelled by all these
$j(\sigma_1)$. In the limit $k\to\infty$, the Hilbert spaces become
infinite-dimensional so that the rule~(S1) in Section~\ref{sect_tqft}
has to be relaxed for the Ponzano--Regge model. The Hilbert spaces
admit a precise interpretation as spaces of states of
$2$-geometries. Although our focus on the partition function instead
of the full path integral with boundary conditions~\eqref{eq_pathint},
seemed to be somewhat narrow at first sight, the examples show that
state sums such as the Turaev--Viro invariant~\eqref{eq_tuvi} already
carry the information about boundary fields and Hilbert
spaces. In fact, once the set of colourings of the state sum has
been specified, it automatically determines the vector spaces and
boundary fields.

Of course, there can be several different formulae of state
sums~\eqref{eq_combinatorial} with different sets of colours which
yield the same invariant. This corresponds to different path
integrals~\eqref{eq_pathint} with different Hilbert spaces that have
the same trace~\eqref{eq_partition}. What we are looking for in the
case of general relativity, is the full TQFT and not just the
partition function. Since all the classification results are available
for manifolds with boundary, the key step is to find the physical
interpretation for the boundary fields, expressed on triangulations of
the boundary, for the set of colours of the state
sum~\eqref{eq_combinatorial}.

\paragraph{Towards 3+1.}

We can now come back to the claim of the introduction that the absence
of gravitons in $d=2+1$ is related
\begin{myenumerate}
\item
  \emph{neither} to the question of whether the path integral
  corresponds to a $C^0$-QFT (as opposed to a $C^\infty$-QFT),
\item
  \emph{nor} to the question of whether the vector spaces of this
  $C^0$-QFT or $C^\infty$-QFT are finite-dimensional,
\item 
  \emph{nor} to the question of whether the theory admits a
  triangulation independent discretization.
\end{myenumerate}
Whereas the question of $C^0$-QFT versus $C^\infty$-QFT is related to
the classification of topological versus smooth manifolds and
therefore depends on the space-time dimension $d$, the question of
finite-dimensionality is presently understood only in $d=2+1$. We have
seen examples for both alternatives, finite-dimensional (Turaev--Viro
model) and infinite-dimensional (Ponzano--Regge model). Both are
constructed from classical theories that have only constant curvature
geometries as their solutions and therefore no propagating modes. The
answer to question (3.) above is finally `yes' in any $d\leq 5+1$ from
the classification results. Concrete examples have so far been
constructed in $d=2+1$ for both pure general relativity (Turaev--Viro
model and Ponzano--Regge model) and for a special case of general
relativity with fermions~\cite{LiOe03}, and in $d=3+1$ only for
BF-theory without~\cite{Oo92} or with~\cite{Ba96,CrKa97} cosmological
constant.

For general relativity in $d=3+1$, the results of
Section~\ref{sect_manifolds} still suggest that we should expect a
Pachner move invariant state sum although the finite-dimensionality of
the Hilbert spaces might be lost in a more drastic fashion than in
$d=2+1$, and there are the remarks of Section~\ref{sect_corners} on
the role of time. A possible approach to $d=3+1$ in order to narrow
down the path integral is suggested by the special properties of
smooth $4$-manifolds as we sketch in the final section.

% ==============================================================================
%
\section{Physics meets Mathematics}
%
% ==============================================================================
\label{sect_mathphys}

Connecting the gauge symmetry of classical general relativity with
results on the classification of smooth manifolds up to
diffeomorphism, we have revealed the coincidence of an open problem in
topology, namely to construct a non-trivial invariant of
piecewise-linear $4$-manifolds, with an open problem in theoretical
physics, namely to find a path integral quantization of general
relativity in $d=3+1$ space-time dimensions.

%------------------------------------------------------------------------------
\subsection{Mathematical aspects}
%------------------------------------------------------------------------------

The mathematical question is whether one can construct invariants of
piecewise-linear $4$-manifolds that are non-trivial in the sense that
they can distinguish inequivalent differentiable structures on the
same underlying topological manifold.

From the classification results, it is known that the Whitehead
triangulation of any given smooth manifold captures the full
information about its differentiable structure up to diffeomorphism
(Appendix~\ref{app_smoothing}). This suggests that a state sum which
probes the abstract combinatorial information contained in a
triangulation, will be sufficient (Appendix~\ref{app_abstract}). On
the other hand, there exist already way too many topological manifolds
in order to extract the complete classification information by any
conceivable algorithm (Appendix~\ref{app_markov}). The interesting
question is therefore whether one can find a suitably restricted class
of piecewise-linear $4$-manifolds, for example closed connected and
simply connected manifolds, for which a non-trivial invariant can be
constructed. The existence of the Donaldson~\cite{Do90} and
Seiberg--Witten~\cite{Wi94} invariants in the smooth framework is
encouraging because it demonstrates that some non-trivial information
can indeed be extracted.

\begin{question}
\label{q_math}
Do there exist state sum invariants of piecewise-linear $4$-manifolds
which are able to distinguish inequivalent PL-structures on the same
underlying topological manifold?
\end{question}

Differential topologists have been considering this question for a
long time, see, for example the introduction of~\cite{FrUh84}. Some
state sum invariants of piecewise-linear $4$-manifolds have already
been constructed, for example the Crane--Yetter
invariant~\cite{CrKa97} and Mackaay's state sum~\cite{Ma99}. So far,
it has not been confirmed that any of these constructions is indeed
non-trivial in the above sense. The Crane--Yetter invariant is known
to depend only on the homotopy type of the underlying topological
manifold~\cite{Ro95} although it offers a novel combinatorial method
for computing the signature (of the intersection form) of
$M$. Mackaay's state sum has so far been explored only for very
special cases in which it, too, depends only on the homotopy
type~\cite{Ma00}. Nevertheless, in order to appreciate this state sum,
a more detailed comparison with the $3$-dimensional case is very
instructive.

Just as the Turaev--Viro invariant~\cite{TuVi92} of $3$-manifolds can
be constructed for a certain class of \emph{spherical
categories}~\cite{BaWe96}, Mackaay's state sum~\cite{Ma99} is defined
for a class of \emph{spherical $2$-categories}. The situations for
which Mackaay's state sum has been carefully studied~\cite{Ma00}
involve rather special spherical $2$-categories for which this state
sum resembles a $4$-dimensional generalization of the
Dijkgraaf--Witten model~\cite{DiWi90}. It is, however, known that the
Dijkgraaf--Witten invariant agrees for lens spaces as soon as they
have the same homotopy type~\cite{AlCo93}. Lens spaces are the
standard examples~\cite{Ro76} of topological $3$-manifolds in order to
show that some invariant can distinguish manifolds of the same
homotopy type that are not homeomorphic.

In order to render the Turaev--Viro invariant non-trivial, the
construction of the (modular categories of representations of the)
quantum groups $U_q(\mathfrak{sl}_2)$, $q=e^{2\pi i/\ell}$,
$\ell\in\N$, seems to be essential. In this case, the Turaev--Viro
invariant indeed distinguishes non-homeomorphic lens spaces of the
same homotopy type\footnote{By the theorem of Turaev and
Walker~\cite{Ro95}, the Turaev--Viro invariant is the squared modulus
of the Reshetikhin--Turaev invariant~\cite{ReTu91} which is
known~\cite{FrGo91} to distinguish, for example, the lens spaces
$L(7,1)$ and $L(7,2)$.}. The same cannot be accomplished with the
category of representations of an ordinary group.

This comparison with the $3$-dimensional case therefore suggests that
Mackaay's state sum should be studied for sufficiently sophisticated
spherical $2$-categories which are as `generic' in the context of
$2$-categories as are the modular categories of representations of
$U_q(\mathfrak{sl}_2)$ in the context of $1$-categories.

%------------------------------------------------------------------------------
\subsection{Physical aspects}
%------------------------------------------------------------------------------

Questions about observables of general relativity are questions about
smooth manifolds and smooth functions. The physical answers are
specified only up to space-time diffeomorphism. Whenever we have made
use of classification results on smooth, piecewise-linear or
topological manifolds, we have exploited this gauge freedom in an
essential way.

The classification results show that unless $d=3+1$, at least the
partition function $Z(M)$~\eqref{eq_partition} of the path integral can
be computed without explicitly referring to the differentiable
structure of the space-time manifold $M$.

In $d=2+1$, $Z(M)$ depends only on the underlying topological manifold
$M$ up to homeomorphism (Appendix~\ref{app_hauptvermutung}), and this
is the mechanism that renders quantum general relativity in $d=2+1$
space-time dimensions particularly simple. This indeed justifies the
jargon `topological'. In $d\leq 1+1$ or $d\geq 4+1$, the information
required in order to determine $Z(M)$ is essentially the homotopy type
of $M$ (assuming that $M$ admits a differentiable structure) together
with information on the tangent bundle and one finite number which
resolves the ambiguities in constructing first a PL-structure for the
topological manifold $M$ (Appendix~\ref{app_hauptvermutung}) and then
a smoothing of this PL-structure (Appendix~\ref{app_smooth}).

Although we have defined the partition function $Z(M)$
in~\eqref{eq_partition} using the path integral of general relativity
in a way that seems to employ the differentiable structure of
space-time, the above results suggest that unless $d=3+1$, there
exists an alternative way of computing $Z(M)$, either from the
underlying topological manifold up to homeomorphism ($d=2+1$), or even
essentially from its homotopy type together with additional choices
($d\leq 1+1$ or $d\geq 4+1$).

It happens only in $d=3+1$, that there can be more than finitely many
pairwise inequivalent differentiable structures for the topological
manifold underlying $M$. This means that there is generically no short
cut available in order to compute $Z(M)$ from the underlying
topological manifold alone. Only in $d=3+1$, differential topology is
rich enough in order to provide us with a large number of non-trivial
$Z(M)$ in the context of smooth manifolds. This is very reassuring
because in $d=3+1$, one might expect quite a number of quantum field
theories with a gauge symmetry under space-time diffeomorphisms even
though the theories other than general relativity are usually treated
in a different framework.

Besides the partition function, will the entire $C^\infty$-QFTs in
space-time dimensions other than $d=3+1$ be necessarily less
interesting? This question cannot be ultimately answered yet, but one
can expect that the dominance of homotopy type in this problem in
$d\leq 1+1$ or $d\geq 4+1$ ($d\geq 5+1$ if there is a non-empty
boundary) would provide strong constraints. Concerning $d=3+1$, we
arrive at,

\begin{question}
Does there exist a particular state sum invariant of piecewise-linear
$4$-mani\-folds, a degenerate limit of which yields the path integral
quantization of pure general relativity in $d=3+1$ space-time
dimensions (in the special case in which no boundary conditions are
imposed)?
\end{question}

Let us again compare the situation with the toy model of general
relativity in $d=2+1$. The state sum that yields the actual
topological invariant, is the Turaev--Viro
invariant~\cite{TuVi92}. This state sum for $U_q(\mathfrak{sl}_2)$,
$q=e^{i\pi/k}$, $k=1,\frac{3}{2},2,\ldots$, corresponds to the
partition function of quantum gravity with Riemannian signature
$\eta=\diag(1,1,1)$ and quantized positive cosmological constant
$\Lambda=4\pi^2/{(\hbar G k)}^2$~\cite{MiTa92,Ba03}. This model
defines~\cite{TuVi92} a proper TQFT in the strict sense~\cite{At88},
based on topological manifolds up to homeomorphism and involving
finitely generated modules. Quantum gravity with Riemannian signature,
but $\Lambda=0$, can then be understood as a limit of this invariant
for $k\to\infty$, $\Lambda\to 0$, $q\to 1$, in which the Hilbert
spaces become infinite-dimensional and the partition function
diverges. The partition function is therefore no longer a well-defined
invariant of topological manifolds. Nevertheless, this degenerate
limit precisely agrees with the original model of
Ponzano--Regge~\cite{PoRe68} which had been invented as a
non-perturbative quantization of general relativity in $d=2+1$ with
Riemannian signature, well before it was realized that this framework
is closely related to invariants of topological manifolds. Quantum
gravity in $d=2+1$ with the realistic Lorentzian signature
$\eta=\diag(-1,1,1)$ is finally even more complicated, replacing
$\Spin(3)$ by $\Spin(2,1)$ and the discrete representation labels of
the Ponzano--Regge model with continuous ones~\cite{Da01,Fr01}.

If we try to extrapolate this experience from $d=2+1$ to $d=3+1$, on
the mathematical side we may well expect that the proper invariant of
piecewise-linear $4$-manifolds involves some highly non-trivial
$2$-category which may be very difficult to guess. Its physical
counterpart, the sought-after path integral of general relativity,
however, needs to be `just' a degenerate limit of the proper
invariant.

Given the experience from $d=2+1$ and noting that the Ponzano--Regge
model (although divergent and therefore not well defined in the
mathematical sense) is so much easier than the Turaev--Viro invariant,
it seems to be a reasonable strategy to approach the state sum
invariant of piecewise-linear $4$-manifolds via theoretical
physics. This means to proceed in two steps: first to construct a
physically motivated, but degenerate limit of the invariant which
corresponds to general relativity; second to study the deformation
theory of the relevant categories and to aim for the actual invariant.

The interesting perspective is here the combination of techniques
developed in mathematics on how to lift combinatorial and algebraic
structures from three to four dimensions, see, for example the
introductory sections of~\cite{Ma99,Ma00} and
also~\cite{BaDo95,CaKa99} for references, with methods from
theoretical physics on how to construct discrete physical models
related to the path integral of general relativity, see, for example,
the review articles~\cite{Ba99,Or01,Pe03}.

The framework outlined here,
\begin{myitemize}
\item
  Does not involve any new physical assumptions. It just combines
  quantum theory (represented by the axioms~(A1), (A2), (A4) of
  axiomatic $X$-QFT for $X=C^0,C^\infty,\PL,h$) with the properties of
  generally relativistic theories (represented by axiom~(A3) and the
  choice $X=C^\infty$).
\item
  Singles out the space-time dimension $d=3+1$.
\item
  Explains that, if mathematicians solve Question~\ref{q_math}, this
  will provide physicists with (a family of) rigorously defined path
  integrals which have the same symmetries as general relativity.
\item
  Explains why the diffeomorphism gauge symmetry takes care of
  renormalization.
\item
  Contains the spin foam models of $(2+1)$-dimensional quantum gravity
  as well-studied examples.
\item
  Applies to other coordinate-free formulated field theories as well
  (\cf\ Section~\ref{sect_diffeo}), not necessarily generally
  relativistic.
\end{myitemize}

In the search for a path integral quantization of general relativity
in $d=3+1$, we are facing a coincidence of open questions in
mathematics with open questions in theoretical physics. In exploring
these connections, we have just scratched the surface.

%------------------------------------------------------------------------------
\acknowledgments
%------------------------------------------------------------------------------

The author is indebted to Marco Mackaay for explaining various results
on the topology of four-manifolds. I would like to thank Louis Crane,
Gary Gibbons, Robert Helling and Daniele Oriti for discussions.

\appendix

% ==============================================================================
%
\section{Topological, piecewise-linear and smooth manifolds}
%
% ==============================================================================

For mathematical background on topological, smooth and
piecewise-linear manifolds, we refer to various
textbooks~\cite{MiTh73,Wa84,Hu69,RoSa72,Mu63} as well as to the
introductory sections of~\cite{FrUh84,GoSt99}. We review here only
those facts that are relevant to the classification of the various
types of manifolds, hoping that this compilation of definitions and
results will make the literature more accessible.

%------------------------------------------------------------------------------
\subsection{Topological manifolds}
%------------------------------------------------------------------------------

We first recall the definitions of topological and smooth manifolds.

\begin{definition}
\label{def_topological}
Let $M$ be a topological space, and fix some $d\in\N$, called the
\emph{dimension}.
\begin{myenumerate}
\item  
  A \emph{chart} or \emph{coordinate system} for $M$ is a pair
  $(U,\phi)$ of an open set $U\subseteq M$ and a homeomorphism
  $\phi\colon U\to\phi(U)$ onto an open subset
  $\phi(U)\subseteq\R^d_+$ of the half-space
  $\R^d_+:=\{x\in\R^d\colon\, x_0\geq 0\}$.
\item
  A $\Top_d$-\emph{atlas} $\mathcal{A}$ for $M$ is a family
  $\mathcal{A}=\{(U_i,\phi_i)\colon\, i\in I\}$ of coordinate systems
  for $M$ such that the $U_i$ form an open cover of $M$,
\begin{equation}
  \bigcup_{i\in I}U_i = M.
\end{equation}
  Here I denotes some index set.
\item
  On the non-empty overlaps $U_{ij}:=U_i\cap U_j\neq\emptyset$,
  $i,j\in I$, there are homeomorphisms,
\begin{equation}
  \phi_{ji}:=\phi_j\circ\phi_i^{-1}|_{\phi_i(U_{ij})}\colon
    \phi_i(U_{ij})\to\phi_j(U_{ij}),
\end{equation}
  which are called \emph{transition functions}.
\item
  The \emph{boundary} $\del M$ of $M$ is the set of all $p\in M$ for
  which $\phi_i(p)\in\R^d_0$, $\R^d_0:=\{x\in\R^d\colon\, x_0=0\}$.
\item
  Two atlases are called \emph{equivalent} if their union is an atlas.
\end{myenumerate}
\end{definition}

These are the most general atlases we are interested in. They give
rise to topological manifolds.

\begin{definition}
\label{def_topmanifold}
Fix some \emph{dimension} $d\in\N$.
\begin{myenumerate}
\item
  Let $M$ be a topological space. A $\Top_d$-structure $[\mathcal{A}]$
  for $M$ is an equivalence class of $\Top_d$-atlases.
\item
  A \emph{topological} $d$-\emph{manifold} $(M,[\mathcal{A}])$ is a
  paracompact Hausdorff space $M$ equipped with a $\Top_d$-structure
  $[\mathcal{A}]$.
\item
  Two topological manifolds $(M,[\mathcal{A}])$, $(N,[\mathcal{B}])$
  are called \emph{equivalent} if $M$ and $N$ are homeomorphic.
\item
  A topological manifold $(M,[\mathcal{A}])$ is called \emph{compact}
  if the underlying topological space $M$ is compact. It is called
  \emph{closed} if it is compact and $\del M=\emptyset$.
\end{myenumerate}
\end{definition}

We often write just $M$ rather than $(M,[\mathcal{A}])$. The boundary
$\del M$ of any topological $d$-manifold $M$ forms a topological
$(d-1)$-manifold with empty boundary. A $0$-manifold is just a set of
points with the discrete topology.

Due to paracompactness, each atlas of any topological manifold admits
a locally finite refinement for which the $\phi_i(U_i)$ are contained
in compact subsets of $\R^d_+$.

The remainder of this section is a technical detail which is necessary
in order to combine results from various different sources in the
literature. The conditions that the underlying topological space in
Definition~\ref{def_topmanifold}(2.) be paracompact and Hausdorff, are
the same as those used in~\cite{Wa84} which are those that correspond
to the physically relevant space-times~\cite{Ge68}. The following
alternative choices are common in the literature,
\begin{myenumerate}
\item
  metrizable~\cite{KiSi69,HiMa74},
\item
  second countable and Hausdorff~\cite{Hu69},
\item
  separable and Hausdorff~\cite{GoSt99}.
\end{myenumerate}

Concerning~(1.), by a theorem of Stone~\cite{HoYo61}, each metrizable
topological space is paracompact and Hausdorff. Conversely, any
paracompact Hausdorff space with a $\Top_d$-atlas inherits a metric
from $\R^d$ on each chart and, by employing a continuous partition of
unity, can be shown to be metrizable.

Concerning~(2.), each topological space with a $\Top_d$-atlas is
locally compact, and any locally compact and second countable
Hausdorff space is metrizable~\cite{HoYo61}. Conversely, by a theorem
of Alexandroff~\cite{HoYo61}, each locally compact metrizable space
admits a countable basis of open sets for each connection
component. As soon as the discussion is restricted to topological
spaces with a countable number of connection components, (2.) is
therefore equivalent to~(1.).

Concerning~(3.), each second countable topological space is separable,
but (3.) is in general weaker than~(2.). We note, however, that a
given topological space is paracompact and locally compact if and only
if it is a free union of spaces that are $\sigma$-compact (unions of
countably many compact sets)~\cite{Du75}. The relevant examples
of~\cite{GoSt99} are all of this type.

%------------------------------------------------------------------------------
\subsection{Smooth manifolds}
%------------------------------------------------------------------------------

We can impose additional structure on manifolds by restricting the
transition functions to appropriate sub-families of
homeomorphisms. Recall that a map $f\colon U\to\R^n$, for some open
set $U\subseteq\R^m$, is called $C^k$ if all $k$-th partial
derivatives exist and are continuous on $U$. A $C^k$-diffeomorphism is
an invertible $C^k$-map with a $C^k$-inverse.

\begin{definition}
Let $M$ be a topological $d$-manifold, $d\in\N$, and
$k\in\N_0\cup\{\infty\}$.
\begin{myenumerate}
\item
  A $C^k$-\emph{atlas} for $M$ is a $\Top_d$-atlas
  $\mathcal{A}=\{(U_i,\phi_i)\colon\, i\in I\}$ such that for all
  $i,j\in I$ with $U_i\cap U_j\neq\emptyset$, the transition functions
  $\phi_{ji}$ are $C^k$-diffeomorphisms.
\item
  A $C^k$-\emph{structure} $[\mathcal{A}]$ for $M$ is an equivalence
  class of $C^k$-atlases.
\end{myenumerate}
\end{definition}

\noindent
This definition includes the topological case
(Definition~\ref{def_topological}) for $k=0$.

\begin{definition}
\label{def_smoothmfd}
Let $k\in\N_0\cup\{\infty\}$.
\begin{myenumerate}
\item
  A \emph{$d$-dimensional $C^k$-manifold} $(M,[\mathcal{A}])$,
  $d\in\N$, is a topological $d$-manifold $M$ equipped with a
  $C^k$-structure $[\mathcal{A}]$.
\item
  Let $(M,[\mathcal{A}])$, $(N,[\mathcal{B}])$ be $C^k$-manifolds with
  atlases $\mathcal{A}=\{(U_i,\phi_i)\colon i\in I\}$ and
  $\mathcal{B}=\{(V_j,\psi_j)\colon j\in J\}$, not necessarily of the
  same dimension. A map $f\colon M\to N$ is called $C^k$ if $f$ is
  continuous and if for all $p\in M$ such that $p\in U_i$, $f(p)\in
  V_j$, for some $i\in I$, $j\in J$, the map,
\begin{equation}
\label{eq_diffcoord}
  \psi_j\circ f\circ\phi_i^{-1}\colon\phi_i(U_i)\to\psi_j(V_j)
\end{equation}
  is a $C^k$-map.
\item
  A \emph{$C^k$-diffeomorphism} $f\colon M\to N$ is an invertible
  $C^k$-map whose inverse is $C^k$.
\item
  Two $C^k$-manifolds are called \emph{equivalent} if they are
  $C^k$-diffeomorphic.
\end{myenumerate}
\end{definition}

The boundary $\del M$ of any $d$-dimensional $C^k$-manifold is a
$(d-1)$-dimensional $C^k$-manifold without boundary. 

\begin{definition}
\begin{myenumerate}
\item
  A $C^k$-atlas or $C^k$-structure, $k\geq 1$, is called
  \emph{oriented} if all transition functions $\phi_{ji}$ are
  orientation preserving, \ie\ if they have positive Jacobi
  determinants.
\item
  An \emph{oriented} $d$-dimensional $C^k$-manifold is a topological
  $d$-manifold with an oriented $C^k$-structure.
\item
  A $C^k$-diffeomorphism is called \emph{orientation preserving} if
  all its coordinate representations~\eqref{eq_diffcoord} are
  orientation preserving.
\end{myenumerate}
\end{definition}

The boundary $\del M$ of any oriented $C^k$-manifold is oriented. We
have defined oriented manifolds only for the cases $C^k$, $k\geq 1$
(although this can also be done for topological manifolds). If we
nevertheless mention oriented topological manifolds in the following,
we do this only if the topological manifold admits some oriented
$C^k$-structure that is unique up to orientation preserving
$C^k$-diffeomorphism.

Obviously, each $C^k$-structure is also a $C^\ell$-structure for any
$0\leq\ell<k$, including the topological case $\ell=0$. As long as we
are only interested in $C^r$-manifolds up to equivalence, this time
excluding the topological case, \ie\ $1\leq r\leq\infty$, we can
restrict ourselves to $C^\infty$-manifolds as the following theorem of
Whitney shows, see, for example~\cite{Mu63}.

\begin{theorem}
Let $M$ be a $C^r$-manifold, $1\leq r\leq\infty$, with some
$C^r$-structure $[\mathcal{A}^{(r)}]$ and some $k$ with
$r\leq k\leq\infty$.
\begin{myenumerate}
\item
  There exists a $C^k$-structure $[\mathcal{A}^{(k)}]$ for $M$.
\item
  The $C^r$-manifolds $(M,[\mathcal{A}^{(r)}])$ and
  $(M,[\mathcal{A}^{(k)}])$ are $C^r$-diffeomorphic.
\item
  Let $[\mathcal{B}^{(k)}]$ denote any other $C^k$-structure for
  $M$. Then $(M,[\mathcal{A}^{(k)}])$ and $(M,[\mathcal{B}^{(k)}])$
  are $C^r$-diffeomorphic.
\end{myenumerate}
\end{theorem}

Of all the $C^\ell$-manifolds, $0\leq\ell\leq\infty$, the relevant
types for the purpose of classification are therefore the topological
(or $C^0$-) and the $C^\infty$-manifolds. A $C^\infty$-structure is
usually called a \emph{differentiable structure}. $C^\infty$-maps and
$d$-dimensional $C^\infty$-manifolds are known as \emph{smooth maps}
and \emph{smooth $d$-manifolds}, and $C^\infty$-diffeomorphisms are
often called just \emph{diffeomorphisms}.

%------------------------------------------------------------------------------
\subsection{Simplices and piecewise-linear manifolds}
%------------------------------------------------------------------------------
\label{app_pl}

In this section, we define another type of manifold by restricting the
transition functions of topological manifolds to \emph{piecewise-linear}
(PL) maps. These are the maps that are compatible with triangulations
of the sets $\phi_i(U_i)\subseteq\R^d$ for the coordinate systems
$(U_i,\phi_i)$. Let us start with the notion of a simplex.

\begin{definition}
Fix some dimension $d\in\N$.
\begin{myenumerate}
\item
  The \emph{convex hull} of some finite set of points
  $A=\{p_0,\ldots,p_k\}\subseteq\R^d$, $k\in\N_0$, is the set
\begin{equation}
  [A] = [p_0,\ldots,p_k]:=\{\,\sum_{j=0}^k p_j\lambda_j\colon\quad
    \lambda_j\geq 0,\quad\sum_{j=0}^k\lambda_j= 1\,\}\subseteq\R^d.
\end{equation}
  We also define $[\emptyset]:=\emptyset$.
\item
  A finite set of points $A=\{p_0,\ldots,p_k\}\subseteq\R^d$,
  $k\in\N$, is called \emph{affine independent} if the set of
  vectors $\{p_1-p_0,\ldots,p_k-p_{k-1}\}$ is linearly
  independent. Sets containing only one point are by definition
  affine independent.
\item
  A $k$-\emph{simplex} in $\R^d$, $k\in\N_0$, is a set
  $\sigma\subseteq\R^d$ of the form $\sigma=[p_0,\ldots,p_k]$ for some
  affine independent set $\{p_0,\ldots,p_k\}\subseteq\R^d$. A
  $(-1)$-simplex is by definition the empty set. 
\item
  Let $\sigma=[A]$ and $\tau=[B]$ be simplices in $\R^d$ so that $A$
  and $B$ are affine independent sets. The simplex $\sigma$ is
  called a \emph{face} of $\tau$ if $A\subseteq B$. In this case we
  write $\sigma\preceq\tau$.
\end{myenumerate}
\end{definition}

Note that the definition of affine independence does not depend on the
numbering of the points. For each simplex $\sigma\subseteq\R^d$, there
is a unique finite set $A\subseteq\R^d$ such that $A$ is affine
independent and $\sigma=[A]$. In this case, we call $\Vert\sigma:=A$
the set of \emph{vertices} of $\sigma$. Each face of a given
$k$-simplex, $k\in\N_0$, is an $\ell$-simplex for some $\ell\leq
k$. Each $0$-simplex has got precisely one face, the empty set, and
$\emptyset\preceq\sigma$ for all simplices $\sigma$. The relation
`$\preceq$' is a partial order on the set of all simplices in $\R^d$.

\begin{definition}
\begin{myenumerate}
\item
  A set $S$ of simplices in $\R^d$ is called \emph{locally finite} if
  each $p\in\R^d$ has got a neighbourhood $U$ such that
  $U\cap\sigma\neq\emptyset$ only for a finite number of simplices
  $\sigma\in S$.
\item
  A \emph{polyhedron} $P\subseteq\R^d$ is a set of the form
\begin{equation}
  |S|:=\bigcup_{\sigma\in S}\sigma,
\end{equation}
  for some locally finite set $S$ of simplices in $\R^d$.
\item
  A \emph{simplicial} complex $K$ in $\R^d$ is a locally finite set of
  simplices in $\R^d$ such that,
\begin{myenumerate}
\item
  whenever $\sigma\in K$ and $\tau\preceq\sigma$ for any simplex
  $\tau\subseteq\R^d$, then also $\tau\in K$, and
\item
  if $\tau,\sigma\in K$ then $\tau\cap\sigma\preceq\tau$ and
  $\tau\cap\sigma\preceq\sigma$.
\end{myenumerate}
\item
  A simplicial complex $K$ is called \emph{finite} if $K$ is a finite
  set.
\item
  If $K$ is a simplicial complex in $\R^d$, we denote by $\Vert
  K:=\{\,p\in\R^d\colon\,\{p\}\in K\,\}$ the set of its
  \emph{vertices}.
\item
  If $K$ is a simplicial complex, the set $|K|$ is called its
  \emph{underlying polyhedron}.
\item
  If a polyhedron $P$ is of the form $P=|K|$ for some simplicial
  complex $K$, then $K$ is called a \emph{triangulation} of $P$.
\end{myenumerate}
\end{definition}

Each polyhedron $P\subseteq\R^d$ has got a triangulation. If $P$ is
compact, then there exists a triangulation of $P$ which is a finite
simplicial complex. We can now define piecewise-linear (PL) maps to be
the maps compatible with the triangulations of polyhedra.

\begin{definition}
Let $P\subseteq\R^m$, $Q\subseteq\R^n$ be polyhedra and $P=|K|$ for
some triangulation $K$.
\begin{myenumerate}
\item
  A map $f\colon P\to Q$ is called \emph{piecewise-linear} (PL) if it
  is continuous and if the restrictions $f|_\sigma$ are affine maps
  for each simplex $\sigma\in K$.
\item
  The map $f$ is called \emph{piecewise differentiable} (PD) if it is
  continuous and if the $f|_\sigma$ are $C^\infty$-maps of maximum
  rank for each $\sigma\in K$.
\end{myenumerate}
\end{definition}

The inverse of any piecewise-linear homeomorphism is also
piecewise-linear. The notion of piecewise differentiability plays a
role when we discuss the compatibility of piecewise-linear and smooth
structures in Appendix~\ref{app_smoothing} below.

\begin{definition}
Let $K$ and $L$ be simplicial complexes. A map $f\colon|K|\to|L|$ is
called \emph{simplicial} is it is continuous, if $f$ maps vertices to
vertices, \ie\ $f(p)\in\Vert L$ for all $p\in\Vert K$, and if for each
$k$-simplex $[p_0,\ldots,p_k]\in K$, the simplex generated by the
images $[f(p_0),\dots,f(p_k)]$ is contained in $L$. A \emph{simplicial
isomorphism} is a simplicial map that is a homeomorphism.
\end{definition}

Let $P$ and $Q$ be polyhedra. A map $f\colon P\to Q$ is PL if and only
if $f$ is a simplicial map for some triangulations $K$ and $L$ such
that $P=|K|$ and $Q=|L|$.

\begin{definition}
\begin{myitemize}
\item
  An \emph{oriented} $k$-simplex $(\sigma,\leq)$ is a $k$-simplex
  $\sigma$ together with a linear order `$\leq$' on the set of its
  vertices $\Vert\sigma$. In the bracket notation $[q_0,\ldots,q_k]$
  for $k$-simplices, we can write for the oriented simplex,
\begin{equation}
  \epsilon\cdot[p_{\tau(0)},\ldots,p_{\tau(k)}],
\end{equation}
  where $\tau\in\mathcal{S}_{k+1}$ is a permutation such that
  $p_0\leq\cdots\leq p_k$ and $\epsilon:=\sgn\tau\in\{-1,+1\}$ is the
  sign of the permutation. For example, for $2$-simplices,
\begin{equation}
  [p_1,p_0,p_2] = -[p_0,p_1,p_2].
\end{equation}
\item
  The faces of an oriented $k$-simplex
  $\sigma=[p_0,\ldots,p_k]\subseteq\R^d$ are the oriented simplices
  with the \emph{induced orientation},
\begin{equation}
  {(-1)}^j\cdot [p_0,\ldots,\hat p_j,\ldots,p_k],
\end{equation}
  where the hat denotes the omission of a vertex.
\item
  An \emph{oriented} simplicial complex $(K,\leq)$ is a simplicial
  complex $K$ together with a partial order `$\leq$' on the set of
  vertices $\Vert K$ that restricts to a linear order on $\Vert\sigma$
  for each $\sigma\in K$.
\item
  A simplicial isomorphism $f\colon|K|\to|L|$ between oriented
  simplicial complexes $(K,\leq)$ and $(L,\leq)$ is called
  \emph{orientation preserving} if it is compatible with the partial
  order, \ie\ if $p\leq q$ implies $f(p)\leq f(q)$ for all
  $p,q\in\Vert K$.
\end{myitemize}
\end{definition}

Each $d$-simplex in $\R^d$, $\sigma=[p_0,\ldots,p_d]$, inherits an
orientation from $\R^d$ such that the linear order of its vertices is
$p_0\leq\cdots\leq p_k$ up to an even permutation precisely when
$\det(p_0-p_1,\ldots,p_{d-1}-p_d) > 0$. Each subset $U\subseteq\R^d_+$
that is open in $\R^d_+$, is contained in some polyhedron which has a
triangulation in terms of an oriented simplicial complex, compatible
with the orientation inherited from $\R^d$.

We are now ready to restrict the transition functions of manifolds to
piecewise-linear maps.

\begin{definition}
Let $M$ be a topological $d$-manifold, $d\in\N$.
\begin{myenumerate}
\item
  A $\PL_d$-\emph{atlas} for $M$ is a $\Top_d$-atlas
  $\mathcal{A}=\{(U_i,\phi_i)\colon\, i\in I\}$ such that for all
  $i,j\in I$ with $U_i\cap U_j\neq\emptyset$, the transition functions
  $\phi_{ji}$ are piecewise-linear homeomorphisms.
\item
  An \emph{oriented} $\PL_d$-atlas for $M$ is a $\PL_d$-atlas such
  that the transition functions $\phi_{ji}$ are orientation preserving
  simplicial maps with respect to the orientation inherited from
  $\R^d$.
\item
  An [oriented] $\PL_d$-\emph{structure} $[\mathcal{A}]$ for $M$ is an
  equivalence class of [oriented] $\PL_d$-atlases.
\end{myenumerate}
\end{definition}

\noindent
The following is as usual.

\begin{definition}
\begin{myenumerate}
\item
  A $d$-dimensional [oriented] \emph{PL-manifold} $(M,[\mathcal{A}])$,
  $d\in\N$, is a topological $d$-manifold $M$ equipped with an
  [oriented] PL-structure $[\mathcal{A}]$.
\item
  Let $(M,[\mathcal{A}])$, $(N,[\mathcal{B}])$ be PL-manifolds with
  atlases $\mathcal{A}=\{(U_i,\phi_i)\colon i\in I\}$ and
  $\mathcal{B}=\{(V_j,\psi_j)\colon j\in J\}$. A map $f\colon M\to N$
  is called \emph{PL} if $f$ is continuous and if for any $p\in M$
  such that $p\in U_i$, $f(p)\in V_j$ for some $i\in I$, $j\in J$, the
  map,
\begin{equation}
\label{eq_plcoord}
  \psi_j\circ f\circ\phi_i^{-1}\colon\phi_i(U_i)\to\psi_j(V_j)
\end{equation}
  is piecewise-linear.
\item
  A PL-isomorphism $f\colon M\to N$ is an invertible PL-map whose
  inverse is PL.
\item
  A PL-isomorphism is called \emph{orientation preserving} if all its
  coordinate representations~\eqref{eq_plcoord} are orientation
  preserving simplicial maps.
\item
  Two PL-manifolds are called \emph{equivalent} if they are
  PL-isomorphic.
\end{myenumerate}
\end{definition}

The boundary $\del M$ of each $d$-dimensional [oriented] PL-manifold
is a $(d-1)$-dimensional [oriented] PL-manifold without boundary.

%------------------------------------------------------------------------------
\subsection{Triangulations of smooth manifolds}
%------------------------------------------------------------------------------
\label{app_smoothing}

A smoothing of some PL-manifold is a differentiable structure that is
compatible with the PL-structure in the following way.

\begin{definition}
Let $M$ be a PL-manifold with the PL-atlas
$\mathcal{A}=\{(V_i,\psi_i)\colon\,i\in I\}.$ A differentiable
structure represented by some $C^\infty$-atlas
$\{(U_j,\phi_j)\colon\,j\in J\}$ on $M$ is called a \emph{smoothing}
of the PL-structure $[\mathcal{A}]$ if on each non-empty overlap
$U_j\cap V_i\neq\emptyset$, $i\in I$, $j\in J$, the homeomorphism
\begin{equation}
  \phi_j\circ\psi_i^{-1}|_{\psi_i(U_j\cap V_i)}\colon
    \psi_i(U_j\cap V_i)\to\phi_j(U_j\cap V_i)
\end{equation}
is piecewise-differentiable.
\end{definition}

Whitehead's theorem\cite{Wh40} guarantees that any smooth manifold of
arbitrary dimension can be triangulated in this way.

\begin{theorem}[Whitehead]
\label{thm_whitehead}
For each smooth manifold $M$, there exists a PL-manifold
$M_\mathrm{PL}$, called its \emph{Whitehead triangulation}, such that
$M$ is diffeomorphic to a smoothing of $M_\mathrm{PL}$. Any two
diffeomorphic smooth manifolds have PL-isomorphic Whitehead
triangulations.
\end{theorem}

%------------------------------------------------------------------------------
\subsection{Combinatorial manifolds}
%------------------------------------------------------------------------------
\label{app_combinatorial}

We have defined PL-manifolds above as topological manifolds subject to
the additional condition that their transition functions are PL. In
this section, we review the concept of combinatorial manifolds which
reduces the study of an entire PL-manifold to the study of a single
simplicial complex and its combinatorics. Notice that the underlying
polyhedron $|K|\subseteq\R^n$ of any simplicial complex $K$ in $\R^n$,
forms a paracompact Hausdorff space with the relative topology induced
from $\R^n$.

\begin{definition}
The underlying polyhedron $|K|$ of a simplicial complex $K$ in $\R^n$
for some $n\in\N$, is called a \emph{combinatorial $d$-manifold} if it
forms a $d$-dimensional PL-manifold, $d\in\N$.
\end{definition}

In order to describe the relationship between PL-manifolds and
combinatorial manifolds, we need the notion of the join and the link
of simplices.

\begin{definition}
Let $\sigma,\tau\subseteq\R^d$ be simplices, $\sigma=[A]$, $\tau=[B]$
with affine independent sets $A,B\subseteq\R^d$.
\begin{myenumerate}
\item
  $\sigma$ and $\tau$ are called \emph{joinable} if the set $A\cup B$
  is affine independent.
\item
  If $\sigma$, $\tau$ are joinable, their \emph{join} is defined to be
  the set $\sigma\cdot\tau:=[A\cup B]$.
\end{myenumerate}
\end{definition}

\begin{definition}
Let $K$ be a simplicial complex and $\sigma\in K$. The \emph{link} of
$\sigma$ in $K$ is the following set of simplices,
\begin{equation}
  \lk_K(\sigma):=\{\tau\in K\colon\quad\mbox{$\sigma$ and $\tau$ are joinable and
    $\sigma\cdot\tau\in K$}\,\}.
\end{equation}
\end{definition}

Notice that $\lk_K(\sigma)$ is itself a simplicial complex. We also
need the definition of PL-balls and PL-spheres.

\begin{definition}
Let $k\in\N$ and $\{p_0,\ldots,p_k\}\subseteq\R^d$ be affine
independent. 
\begin{myenumerate}
\item
  A piecewise-linear $k$-\emph{ball} is a polyhedron which is
  PL-isomorphic to the $k$-simplex,
\begin{equation}
  B^k:=[p_0,\ldots,p_k].
\end{equation}
\item
  For each simplex $\sigma$, we define the simplicial complex
  $\overline\sigma:=\{\tau\colon\,\tau\preceq\sigma\}$ which contains
  $\sigma$ together with all its faces.
\item
  A piecewise-linear $(k-1)$-\emph{sphere} is a polyhedron which is
  PL-isomorphic to the polyhedron,
\begin{equation}
  S^{k-1}=\del B^k:=|\{\sigma\in\overline{B^k}\colon\,\sigma\neq B^k\}|.
\end{equation}
\end{myenumerate}
\end{definition}

A $0$-ball is a set containing one point, and a $0$-sphere a set with
precisely two points. $\ell$-balls and $\ell$-spheres for $\ell<0$ are
by definition the empty set. PL-manifolds can be characterized by
combinatorial manifolds as follows.

\begin{theorem}
\label{thm_combinatorial}
Each PL-manifold is PL-isomorphic to some combinatorial
manifold. Conversely, the underlying polyhedron $|K|\subseteq\R^n$ of
some simplicial complex $K$ admits a $\PL_d$-structure, $n,d\in\N$, if
and only if for each $k$-simplex $\sigma\in K$, the polyhedron
$|\lk_K(\sigma)|$ is a PL $(d-k-1)$-ball or a PL $(d-k-1)$-sphere.
\end{theorem}

This theorem implies that each PL-manifold $M$ can be described by a
single simplicial complex $K$ with certain properties. If $M$ is
compact, then $K$ can be chosen to be finite. Otherwise, due to
paracompactness, $K$ can always be built up from only a countable
number of simplices. Furthermore, $\sigma\in K$ is contained in the
boundary $\del M$ if and only if $|\lk_K(\sigma)|$ is a PL
$(d-k-1)$-ball (as opposed to a sphere). Each $(d-1)$-simplex in $K$
is the face of at most two $d$-simplices in which it occurs with
opposite induced orientations.

%------------------------------------------------------------------------------
\subsection{Abstract triangulations}
%------------------------------------------------------------------------------
\label{app_abstract}

It is finally even possible to forget the surrounding standard space
$\R^n$ in which the simplices are contained, and to concentrate only
on their combinatorics.

\begin{definition}
An \emph{abstract simplicial complex} $(K^0,K)$ is a set $K^0$, called
the set of \emph{abstract vertices}, together with a family
$K\subseteq\mathcal{P}(K^0)$ of finite subsets of $K^0$, called the
set of \emph{abstract simplices}, such that
\begin{myenumerate}
\item
  $\{p\}\in K$ for all $p\in K^0$, and
\item
  If $\sigma\in K$ and $\tau\subset\sigma$, then also $\tau\in K$.
\end{myenumerate}
For some subset $A\subseteq K$ of abstract simplices, we denote by
$(\overline{A}^0,\overline{A})$ the smallest abstract simplicial
complex that contains all simplices of $A$, \ie\
$\overline{A}=\{\,\sigma\in K\colon\, \sigma\subseteq\tau, \tau\in
A\,\}$ and $\overline{A}^0=\{\,p\in K^0\colon\, \{p\}\in A\,\}$. We
often write just $K$ rather than $(K^0,K)$. An \emph{oriented abstract
simplicial complex} is an abstract simplicial complex with a partial
order on the set $K^0$ that restricts to a linear order on each
$\sigma\in K$.
\end{definition}

Given some simplicial complex $L$ in $\R^n$, its \emph{abstraction} is
the abstract simplicial complex $(K^0,K)$ for which $K^0=\Vert L$ and
$\{p_0,\ldots,p_k\}\in K$ if and only if $[p_0,\ldots,p_k]\in
L$. Conversely, given some finite abstract simplicial complex
$(K^0,K)$, there exists a simplicial complex $L$ in $\R^{|K^0|-1}$
whose abstraction is $(K^0,K)$. $L$ is unique up to simplicial
isomorphism. If $(K^0,K)$ is a finite abstract simplicial complex, we
define its \emph{underlying polyhedron} as $|K|:=|L|$ which is well
defined up to simplicial isomorphism. Recall the additional conditions
(Theorem~\ref{thm_combinatorial}) if this is supposed to yield a
PL-manifold.

We can now study the question of whether any two given PL-manifolds
are PL-isomorphic in the context of the abstractions of their global
triangulations. This is accomplished in a systematic way for any
dimension by Pachner's theorem~\cite{Pa91}. We mention here the
version for closed manifolds.

\begin{theorem}[Pachner]
Let $K$ and $L$ be the finite abstract simplicial complexes associated
with closed oriented combinatorial $d$-manifolds $|K|$ and $|L|$,
$d\in\N$.  The combinatorial manifolds $|K|$ and $|L|$ are
PL-isomorphic if and only if the abstract simplicial complexes $K$ and
$L$ are related by a finite sequence of bistellar moves, the so-called
\emph{Pachner moves}.
\end{theorem}

The Pachner moves in dimension $d\in\N$ can be described as
follows. Consider an oriented $(d+1)$-simplex
$\sigma=[0,1,\ldots,d+1]$ where we write integer numbers for the
linearly ordered vertices. Write down the set of faces of
$\sigma$. These are $d+2$ oriented $d$-simplices,
\begin{equation}
  \{\,+[1,2,\ldots,d+1],-[0,2,3,\ldots,d+1],+[0,1,3,4,\ldots,d+1],\ldots,{(-1)}^d\,[0,1,\ldots,d]\,\}.
\end{equation}
For each $\ell\in\N$, $1\leq\ell\leq d/2+1$, partition this set into
the following subsets,
\begin{mathletters}
\begin{eqnarray}
  A_\ell&=&\{\,+[1,2,\ldots,d+1],\ldots,{(-1)}^{\ell-1}\,[0,\ldots,\hat{\ell-1},\ldots,d+1]\,\},\\
  B_\ell&=&\{\,-{(-1)}^\ell\,[0,\ldots,\hat\ell,\ldots,d+1],\ldots,-{(-1)}^d\,[0,1,\ldots,d]\,\},
\end{eqnarray}%
\end{mathletters}%
where the hat denotes omission of a vertex and where we have reversed
the orientation of all simplices in $B_\ell$.

Both $A_\ell$ and $B_\ell$ generate oriented abstract simplicial
complexes $\overline{A_\ell}$ and $\overline{B_\ell}$. Observe that
both polyhedra $|\overline{A_\ell}|$ and $|\overline{B_\ell}|$ have
PL-isomorphic boundaries. The \emph{bistellar $\ell$-move} consists of
cutting out the simplices of $A_\ell$ from the given abstract
simplicial complex and gluing in those of $B_\ell$. The Pachner moves
in dimension $d$ are the bistellar $\ell$-moves, $1\leq\ell\leq
d/2+1$, and their inverses. For manifolds with boundary, the Pachner
moves are the so-called \emph{elementary shellings}~\cite{Pa91}.

%------------------------------------------------------------------------------
\subsection{Homotopy type}
%------------------------------------------------------------------------------
\label{app_homotopytype}

Topological manifolds can not only be compared by studying whether
they are homeomorphic or not, but there is also the weaker notion of
\emph{homotopy type}.

\begin{definition}
Let $M$, $N$ be topological spaces. 
\begin{myenumerate}
\item
  Continuous maps $f,g\colon M\to N$ are called \emph{homotopic}
  ($f\simeq g$) if there exists a continuous map $F\colon[0,1]\times
  M\to N$ such that $F(0,p)=f(p)$ and $F(1,p)=g(p)$ for all $p\in
  M$. The map $F$ is called a \emph{homotopy}.
\item
  $M$ and $N$ are called \emph{homotopy equivalent} or \emph{of the
  same homotopy type} ($M\simeq N$) if there exist continuous maps
  $f\colon M\to N$ and $h\colon N\to M$ such that $f\circ
  h\simeq\id_N$ and $h\circ f\simeq\id_M$.
\end{myenumerate}
\end{definition}

%------------------------------------------------------------------------------
\subsection{Comparison of structures}
%------------------------------------------------------------------------------
\label{app_comparison}

By Theorem~\ref{thm_whitehead}, any two diffeomorphic smooth manifolds
have got PL-isomorphic PL-manifolds as their Whitehead
triangulations. Each PL-manifold has got some underlying topological
manifold, and PL-isomorphic PL-manifolds obviously have homeomorphic
underlying topological manifolds. Finally, homeomorphic topological
manifolds are always of the same homotopy type. We therefore have the
following hierarchy,
\begin{equation}
\label{eq_classification}
  \mbox{diffeomorphic}\Longrightarrow\mbox{PL-isomorphic}\Longrightarrow
  \mbox{homeomorphic}\Longrightarrow\mbox{homotopy\ equivalent}
\end{equation}
Which of these arrows can be reversed? It turns out that the answer to
this question strongly depends on the dimension. Before we comment of
the individual arrows in greater detail, we mention the following
algebraic limitation to the classification of manifolds.

%- - - - - - - - - - - - - - - - - - - - - - - - - - - - - - - - - - - - - - -
\subsubsection{Impossibility of complete classification}
%- - - - - - - - - - - - - - - - - - - - - - - - - - - - - - - - - - - - - - -
\label{app_markov}

A complete classification of manifolds of dimension $d\geq 4$ is not
possible, even if the question is restricted to the classification of
topological manifolds up to homotopy equivalence. Manifolds $M\simeq
N$ of the same homotopy type have isomorphic fundamental groups
$\pi_1(M)\cong\pi_1(N)$. The problem is that for any $d\geq 4$ and for
any group $G$ given by a finite presentation in terms of generators
and relations, there exists a closed topological $d$-manifold $M$ with
$\pi_1(M)=G$, see, for example~\cite{GoSt99}. But there exists no
algorithm that decides for any two given finite group presentations
whether they describe isomorphic groups or not~\cite{Ra58}.

This algebraic difficulty is the reason why some results on
$d$-manifolds, $d\geq 4$, are usually stated only for closed,
connected and simply connected manifolds. In the case of some generic
closed and connected manifold $M$, one would therefore study its
universal covering $\hat M$ first, but then the covering map itself
whose fibres are just $\pi_1(M)$, can be extremely complicated.

%- - - - - - - - - - - - - - - - - - - - - - - - - - - - - - - - - - - - - - -
\subsubsection{PL-isomorphism versus diffeomorphism}
%- - - - - - - - - - - - - - - - - - - - - - - - - - - - - - - - - - - - - - -
\label{app_smooth}

In dimensions $d\leq 6$, the equivalence classes of smooth manifolds
up to diffeomorphism are in one-to-one correspondence with those of
PL-manifolds up to PL-isomorphism as the following theorem shows.

\begin{theorem}
\label{thm_smoothing}
Let $d\leq 6$. Each $d$-dimensional PL-manifold admits a smoothing,
and any two PL-isomorphic PL-manifolds of dimension $d$ give rise to
diffeomorphic smoothings.
\end{theorem}

For $d\leq 3$, this theorem is a classical result which already holds
for the underlying topological manifolds and the unique smooth
structure they admit~\cite{Mo53}. For $d\in\{5,6\}$, it follows from a
theorem of Kirby--Siebenmann~\cite{KiSi69} whereas in $d=4$, it is a
consequence of a result by Cerf~\cite{Ce68} in the context of the
general smoothing theory, see, for example~\cite{Ku62,HiMa74}.

For a given $d$-dimensional PL-manifold, $d\geq 7$, there can be
obstructions to the existence of smoothings as well as a finite
ambiguity, \ie\ there can exist a finite number of pairwise
non-diffeomorphic smoothings, see, for example~\cite{KiSi69}. An
example for such an ambiguity is provided by the exotic smooth
structures of $S^7$~\cite{KeMi63}, and there exists an $8$-dimensional
PL-manifold that does not admit any differentiable structure at
all~\cite{KiSi69}. Usually, the smoothings of some given PL-manifold
(and also the PL-structures of some given topological manifold) are
classified not up to diffeomorphism, but up to the stronger relation
of isotopy~\cite{KiSi69,HiMa74}. We have formulated the results above
(Theorem~\ref{thm_whitehead} and~\ref{thm_smoothing}) in such a way
that this subtlety is hidden: If two $d$-dimensional PL-manifolds,
$d\leq 6$, are PL-isomorphic, then their smoothings are
diffeomorphic. The corresponding diffeomorphism, however, is in
general not the same map as the PL-isomorphism with which one has
started, but only isotopic to it.

%- - - - - - - - - - - - - - - - - - - - - - - - - - - - - - - - - - - - - - -
\subsubsection{Homeomorphism versus PL-isomorphism}
%- - - - - - - - - - - - - - - - - - - - - - - - - - - - - - - - - - - - - - -
\label{app_hauptvermutung}

In dimension $d\leq 3$, the equivalence classes of PL-manifolds up to
PL-isomorphism are in one-to-one correspondence with those of
topological manifolds up to homeomorphism because of the following
theorem~\cite{Mo53}.

\begin{theorem}[Bing, Moise]
Let $d\leq 3$. Each topological $d$-manifold admits a PL-structure
(proof of the triangulation conjecture). Homeomorphic topological
$d$-manifolds give rise to PL-iso\-mor\-phic PL-structures (proof of
the Hauptvermutung of Steinitz).
\end{theorem}

This is no longer true if $d=4$. The work of Freedman~\cite{Fr82} and
Donaldson~\cite{Do90} has lead to the construction of families of
smooth manifolds that are homeomorphic, but pairwise
non-diffeomorphic. There exists, for example, an uncountable family of
pairwise non-diffeomorphic differentiable structures on the
topological manifold $\R^4$~\cite{Ta87,GoSt99}, and there are
countably infinite families of homeomorphic, but pairwise
non-diffeomorphic compact smooth
manifolds~\cite{FrMo88,OkVe86}. Furthermore, there exist topological
$4$-manifolds that do not admit any differentiable structure at
all~\cite{Fr82}.

Together with Theorem~\ref{thm_whitehead} and
Theorem~\ref{thm_smoothing}, this implies that in $d=4$, there is a
substantial discrepancy between the classification of topological
manifolds up to homeomorphism and that of PL-manifolds up to
PL-isomorphism.

In dimension $d\geq 5$, there can also be both obstructions to the
existence of a PL-structure on some given topological manifold as well
as an ambiguity, but the ambiguity leaves only a finite number of
choices, see, for example~\cite{KiSi69}. Special attention has to be
paid to $5$-manifolds whose boundary is non-empty because of the
special features in $d=4$.

%- - - - - - - - - - - - - - - - - - - - - - - - - - - - - - - - - - - - - - -
\subsubsection{Homotopy equivalence versus homeomorphism}
%- - - - - - - - - - - - - - - - - - - - - - - - - - - - - - - - - - - - - - -
\label{app_homotopy}

It is a classical result that any two closed oriented topological
$2$-manifolds that are homotopy equivalent, are in fact
homeomorphic. The only invariant required in order to determine the
homotopy type of a closed oriented topological $2$-manifold, is the
genus.

This correspondence disappears in dimension $d=3$ in view of the
\emph{lens spaces} $L(p,q)$. There exist closed oriented topological
$3$-manifolds that are homotopy equivalent, but not homeomorphic, for
example $L(7,1)$ and $L(7,2)$~\cite{Ro76}.

In dimension $d\geq 4$, however, homotopy type is again very important
in the classification of topological manifolds. Very strong results
are provided by the $h$-cobordism theorems of Smale~\cite{Sm62}
($d\geq 5$, even in the smooth case) and of Freedman~\cite{Fr82}
($d=4$, restricted to the topological case). In particular, there is
Freedman's theorem~\cite{Fr82}.

\begin{theorem}[Freedman]
Let $M$, $N$ be connected and simply connected closed topological
$4$-manifolds. If $M$ and $N$ are homotopy equivalent, then they are
homeomorphic.
\end{theorem}

Furthermore, the generalized Poincar{\'e} conjecture is true in
$d=4$~\cite{Fr82} as well as in $d\geq 5$~\cite{Sm62,Ne66}.

\begin{theorem}[generalized Poincar{\'e} conjecture]
Let $M$ be a topological $d$-manifold, $d\geq 4$. If $M$ is homotopy
equivalent to $S^d$, then $M$ is homeomorphic to $S^d$.
\end{theorem}

If one wants to study smooth manifolds and start from the underlying
manifold up to homotopy type, one has to supply additional data on the
structure of the tangent bundle and then resolve a finite ambiguity,
see, for example~\cite{Su77}.

%- - - - - - - - - - - - - - - - - - - - - - - - - - - - - - - - - - - - - - -
\subsubsection{Summary (Table~\ref{tab_classification} of Section~\ref{sect_manifolds})}
%- - - - - - - - - - - - - - - - - - - - - - - - - - - - - - - - - - - - - - -

In dimension $d=2$, all four arrows of~\eqref{eq_classification} can
be reversed, at least for closed oriented manifolds.

In dimension $d=3$, each topological manifold still admits a
differentiable structure that is unique up to diffeomorphism, but
homotopy equivalence no longer implies homeomorphism. 

In dimension $d=4$, each PL-manifold still admits a smoothing that is
unique up to diffeomorphism. Topological manifolds of the same
homotopy type are also homeomorphic, at least for closed, connected
and simply connected manifolds, but homeomorphism no longer implies
diffeomorphism. There can rather exist infinitely many pairwise
inequivalent differentiable structures for the same underlying
topological manifold.

In dimension $d\geq 5$, there can finally be obstructions and
ambiguities both for having PL-structures on a given topological
manifold and for smoothing a given PL-manifold. All the ambiguities in
$d\geq 5$, however, are finite~\cite{KiSi69}. What we have summarized
so far, finally implies Theorem~\ref{thm_intro} of the Introduction.

There are therefore two very striking `gaps' in which the material we
have reviewed so far, does not suffice in order to reverse the arrows
of~\eqref{eq_classification},
\begin{myitemize}
\item
  in $d=3$ to study topological manifolds that are homotopy equivalent
  but not homeomorphic,
\item
  in $d=4$ to study smooth manifolds that are homeomorphic, but not
  diffeomorphic.
\end{myitemize}
The Turaev--Viro invariant~\cite{TuVi92}, closely related to the
partition function $Z(M)$ of quantum gravity in $d=2+1$, provides
non-trivial topological information on the first of these
questions. We speculate that quantum gravity in $d=3+1$ will provide
new insight into the second question.

\end{document}